\documentclass{aastex62}

\graphicspath{{./}{figures/}}

\usepackage[fleqn]{amsmath} 
\usepackage{amssymb}  
\usepackage{gensymb}
\usepackage[nolist]{acronym}
\usepackage{ulem}
\usepackage{bm}
\usepackage{etoolbox}
\usepackage{hyperref}
\usepackage{enumitem}

\definecolor{ochre}{rgb}{0.8, 0.47, 0.13}

\definecolor{CarlGreen}{rgb}{0.30980392, 0.65098039, 0.29803922}

\newcommand{\customlabel}[2]{%
   \protected@write \@auxout {}{\string \newlabel {#1}{{#2}{\thepage}{#2}{#1}{}} }%
   \hypertarget{#1}{}
}

\acrodef{BH}{black hole}
\acrodefplural{BH}[BHs]{black holes}
\acrodef{BBH}{binary black hole}
\acrodefplural{BBH}[BBHs]{binary black holes}
\acrodef{CHE}{chemically homogeneous evolution}
\acrodef{PISN}{pair-instability supernova}
\acrodefplural{PISN}[PISNe]{pair-instability supernova}
\acrodef{IMRI}{intermediate mass-ratio inspiral}
\acrodefplural{IMRI}[IMRIs]{intermediate mass-ratio inspirals}
\acrodef{CE}{common--envelope}
\acrodefplural{CE}[CEs]{common--envelopes }
\acrodef{CEE}{common-envelope episode}
\acrodefplural{CEE}[CEEs]{common--envelope episodes}
\acrodef{RLOF}{Roche-lobe overflow}
\acrodef{IMRAC}{intermediate mass-ratio coalescence}
\acrodefplural{IMRAC}[IMRACs]{intermediate mass-ratio coalescences}
\acrodef{LVC}{LIGO-Virgo Collaboration}
\acrodef{GW}{gravitational-wave}
\newcommand{\event}{GW170729}
\newcommand{\eventGap}{GW190521}

\usepackage{import}

\usepackage{xspace}

\newcommand{\Msun}{\ensuremath{\,\rm{M}_{\odot}}\xspace}

\shorttitle{BBH Mergers from Stellar Triples}
\shortauthors{Vigna-G\'omez et al.}

\begin{document}

\title{Massive Stellar Triples Leading to Sequential Binary Black-Hole Mergers in the Field}

\correspondingauthor{Alejandro Vigna-G\'omez}
\email{avignagomez@nbi.ku.dk}

\author[0000-0003-1817-3586]{Alejandro Vigna-G\'omez}
\affil{DARK, Niels Bohr Institute,
University of Copenhagen,
Jagtvej 128, 2200,
Copenhagen, Denmark}

\author[0000-0002-2998-7940]{Silvia Toonen}
\affil{Institute of Gravitational Wave Astronomy, 
School of Physics and Astronomy, 
University of Birmingham, 
Birmingham, B15 2TT,
United Kingdom}

\author[0000-0003-2558-3102]{Enrico Ramirez-Ruiz}
\affiliation{Department of Astronomy and Astrophysics,
University of California,
Santa Cruz, CA 95064, USA}
\affiliation{DARK, Niels Bohr Institute,
University of Copenhagen,
Jagtvej 128, 2200,
Copenhagen, Denmark}

\author{Nathan W.C. Leigh}
\affiliation{Departamento de Astronom\'ia, 
Facultad de Ciencias F\'isicas y Matem\'aticas, 
Universidad de Concepci\'on, 
Concepci\'on, Chile}
\affiliation{Department of Astrophysics, 
American Museum of Natural History, 
New York, NY 10024, USA}

\author{Jeff Riley}
\affiliation{Monash Centre for Astrophysics, 
School of Physics and Astronomy, 
Monash University, 
Clayton, Victoria 3800, Australia}
\affiliation{The ARC Centre of Excellence for Gravitational Wave Discovery -- OzGrav}

\author[0000-0001-8040-9807]{Carl-Johan Haster}
\affiliation{LIGO Laboratory, 
Massachusetts Institute of Technology, 
185 Albany St, Cambridge, MA 02139, USA}
\affiliation{Department of Physics and Kavli Institute for Astrophysics and Space Research,
Massachusetts Institute of Technology, 
77 Massachusetts Ave, Cambridge, MA 02139, USA}



\begin{abstract}
Stellar triples with massive stellar components are common, and  can lead to sequential binary black-hole mergers. 
Here, we outline the evolution towards these sequential mergers, and explore these events in the context of gravitational-wave astronomy and  the pair-instability mass gap.
We find that binary black-hole mergers in the pair-instability mass gap can be of triple origin and therefore are not exclusively formed in dense dynamical environments.
We discuss the sequential merger scenario in the context of the most massive gravitational-wave sources detected to date: \event\ and \eventGap.
We propose that the progenitor of \event\ is a low-metallicity field triple.
We support the premise that \eventGap\ could not have been formed in the field.
We conclude that triple stellar evolution is fundamental in the understanding of gravitational-wave sources, and likely, other energetic transients as well.
\end{abstract}




\section{Introduction}
\label{sec:intro}
The importance of interactions between massive stars in isolated binaries has become increasingly recognized in the last decades \citep[e.g.,][]{Podsiadlowski1992,Sana2012}. 
Recent studies indicate that early B and O type stars are almost exclusive part of 
higher-order configurations, such as triples and quadruple systems \citep{Sana_2014,MoeDiStefano2017}. 
If future surveys confirm this, our understanding of massive stellar evolution will have to include the increased complexity of multiple-body interactions that were previously mostly considered in dense dynamical environments such as nuclear, globular, or open \textit{clusters} \citep[e.g.,][]{SigurdssonHernquist1993,LeighGeller2013}.

An alternative to electromagnetic methods to study high-mass stellar multiplicity is \ac{GW} astronomy, as massive stars are believed to be the progenitors of stellar-mass \acp{BH}. 
Stellar-mass \acp{BBH} are believed to form predominantly in field binaries \cite[e.g.,][]{belczynski2002comprehensive,Neijssel2019MSSFR}
and clusters \citep[e.g.,][]{PortegiesZwartMcMillan2000,Rodriguez2019Gap,Leigh2014}.  
There have been efforts in trying to understand how best to segregate these two different origins, mostly based on eccentricities and spins, yet there is no definitive consensus on the origin of current \ac{GW} sources \citep[e.g.][]{GWTC12019}.

One candidate signature for cluster origin of a \ac{GW} is a \ac{BH} mass in the \ac{PISN} \textit{mass gap}\footnote{In this {\it Letter} we do not consider the potential mass gap between massive neutron stars and low-mass black holes.}.
\acp{PISN} are initiated by an electron-positron pair-instability which eventually leads to explosive oxygen burning in the core of massive stars \citep[see][and references therein]{Langer2012}.
\acp{PISN} do not leave behind remnants and therefore  a gap is expected in the mass distribution of \acp{BH} for stars with helium core masses in the range of $\approx 64-133\ \rm{M_{\odot}}$ \citep{HegerWoosley2002}. 
Consequently, isolated binary evolution theory does not predict individual \acp{BH} in that regime \citep{Stevenson2019PISN,vanSon2020massGap}. As multiple \ac{BH} mergers can populate the mass gap, the discovery of \ac{BBH} systems with one component lying within the mass gap is considered so far a smoking gun for cluster origin  \citep{Rodriguez2019Gap,SamsingHotokezaka2020}.
The recent detection of \eventGap\ \citep{GW190521PRL}, with at least one \ac{BH} within the mass gap, adds to the conundrum.

Here, we give an overview of which isolated massive stellar triples experience a \textit{sequential merger} of \acp{BBH}. 
We investigate the potential origin of such configurations, put them in the context of \ac{GW} observations, and focus on the masses and spins of sequential mergers leading to \acp{BBH} in the mass gap.
We propose that \event\ is of isolated triple origin, and suggest the mass gap event \eventGap\ was not formed in the field.
Finally, we highlight the importance of massive stellar triples in a broader astronomical context.

\section{Method} 
\label{sec:method}
Here, we outline our main assumptions for the key physics of the formation of a sequential merger starting at the zero-age main sequence.
We consider isolated hierarchical triple systems, composed of an inner binary with masses $M_1 \geq M_2$ and an outer companion of mass $M_3$. 
In the sequential mergers that we consider here the inner \ac{BBH} merges first, and afterwards the remnant merges with the outer \ac{BH}.
We adopt circular coplanar prograde orbits, as supported by observations of compact triples \citep{Tokovinin2017Triple}. 
For circular coplanar prograde orbits we do not expect Lidov-Kozai cycles and neglect other three-body dynamical effects \citep[see, e.g.,][for a review]{Naoz2016}.

The triple must remain dynamically stable from the zero-age main sequence until the inner \ac{BBH} merger, which holds if \citep{MardlingAarseth2001}:
\begin{equation}
\label{eq:triple}
    \dfrac{a_{\rm{out}}}{a_{\rm{in}}} \ge \Bigg( \dfrac{a_{\rm{out}}}{a_{\rm{in}}}\Bigg)_{\rm{crit}} \equiv \dfrac{2.8}{1-e_{\rm{out}}} \Bigg[ \dfrac{(1+q_{\rm{out}})(1+e_{\rm{out}})}{\sqrt{1-e_{\rm{out}}}} \Bigg]^{2/5},
\end{equation}
where $a$ is the semi-major axis, $e$ is the eccentricity and $q$ is the mass ratio (the inner and outer orbits are specified by the {\it in} and {\it out} subscripts). 
The outer mass ratio is $q_{\rm{out}} \equiv M_3/(M_1+M_2)$. 
    
We use the synthetic \ac{BBH} population from \cite{Riley2020CHE} based on isolated binary evolution to investigate the orbital properties of the inner binary (Appendix for further details). 
The models include mass loss, mass transfer, supernovae and \ac{CHE}.
In Figure \ref{fig:totalMass} we present their intrinsic mass distribution of merging \acp{BBH}.
We focus on low-metallicity stars ($Z\lesssim 10^{-3}$) in order to neglect mass loss, spin-down and orbital changes due to stellar winds.
 
As each star of the triple evolves, it will eventually become a \ac{BH} with mass $M_{\rm{BH,i}}$ (with $i=1,2,3$), for which we assume the following.
\acp{BH} have a minimum mass of $M_{\rm{BH,min}}=2.5\ \rm{M_{\odot}}$. 
There is a mass gap between $43 \lesssim M_{\rm{gap}}/\rm{M_{\odot}} \lesssim 124$ \citep{duBuisson2020CHE} due to \ac{PISN} . 
The exact lower limit of the mass gap is  uncertain, and might be as high as $M_{\rm{BH}} \approx 50\ \rm{M_{\odot}}$ \citep[for an overview see][and references therein]{Stevenson2019PISN}.
Furthermore, we assume that stars with carbon-oxygen core masses above $11\ \rm{M_{\odot}}$ experience complete fallback \citep{fryer2012compact} and negligible neutrino mass loss \citep{Muller2016}, likely suppressing \ac{BH} natal kicks.
All \acp{BH} are born as slow rotators \citep{FullerLinhao2019}.
   
During the inner \ac{BBH} merger, mass is lost by radiation from the center of mass of the merging \ac{BBH}.
This leads to a \ac{GW} Blaauw-kick similar to that in spherically symmetric supernovae \citep{blaauw1961origin}. The fraction of radiated mass with respect to the mass of the merging \ac{BBH} ($f_{\rm{rad}}$) depends on the masses and spins of the system (Appendix). 
This modifies the orbit of the tertiary:
\begin{equation}
\label{eq:BlaauwSeparation}
    \dfrac{a_{\rm{out,post}}}{a_{\rm{out,pre}}}=\Bigg[ 2-\dfrac{M_{\rm{BH,1}}+M_{\rm{BH,2}}+M_{\rm{BH,3}}}{M_{\rm{BBH,in}}+M_{\rm{BH,3}}} \Bigg]^{-1},
\end{equation}
and
\begin{equation}
\label{eq:BlaauwEccentricity}
    e_{\rm{out,post}} = \dfrac{f_{\rm{rad}}(M_{\rm{BH,1}}+M_{\rm{BH,2}})}{M_{\rm{BBH,in}}+M_{\rm{BH,3}}},
\end{equation}
where $M_{\rm{BBH,in}}\approx (1-f_{\rm{rad}}) (M_{\rm{BH,1}}+M_{\rm{BH,2}})$.
For non-spinning \acp{BBH} and assuming $e_{\rm{out,pre}}\approx 0$, $f_{\rm{rad}}\approx 0.05$, $e_{\rm{out,post}}\approx 0.05$ and $a_{\rm{out,post}} \approx 1.06 \times a_{\rm{out,pre}}$. 
Furthermore, conservation of momentum gives rise to a recoil kick, which has a magnitude of zero when $q_{\rm{BBH,in}} \approx 0$ or $q_{\rm{BBH,in}} \approx 1$ and can be as high as $v_{\rm{recoil}} \lesssim 175\ \rm{km\ s^{-1}}$ for $q_{\rm{BBH,in}} \approx 0.36$ \citep{GonzalezSperhake2007}. 
As our synthetic population favors $q_{\rm{BBH,in}} \gtrsim 0.9$ (Figures \ref{fig:totalMass} and \ref{fig:massGap}), the magnitude of the recoil kick should be small and we ignore it.
The radiated mass fraction is maximal ($f_{\rm{rad,max}}\approx 0.12$) when we consider a merger from an equal-mass maximally-spinning \ac{BBH} that is aligned with the orbital spin.
This constrains the post-merger eccentricity and separation to $e_{\rm{out}} \lesssim 0.14$ and $a_{\rm{out,post}} \lesssim 1.16 \times a_{\rm{out,pre}}$ respectively.
The systems we consider here can therefore not be unbound during the inner \ac{BBH} merger.

We assume the outer orbit is almost circular at all times.
In this case the \ac{GW} inspiral time can be approximated with \citep{peters1964gravitational}:   
\begin{equation}\label{eq:Peters}
    T_c \approx \dfrac{5}{256} \dfrac{c^5a_{\rm{out}}^4}{G^3M_{\rm{BBH,in}}M_{\rm{BH,3}}(M_{\rm{BBH,in}}+M_{\rm{BH,3}})}.
\end{equation}
    
We estimate the effective spin of the sequential merger as
    \begin{equation}
    \label{eq:spins}
    \chi_{\rm{eff}} = \dfrac{ (M_{\rm{BBH,in}}\vec{\chi}_{\rm{BBH,in}}+M_{\rm{BH,3}}\vec{\chi}_{\rm{BH,3}}) \cdot \hat{L}_{\rm{N}} }{M_{\rm{BBH,in}}+M_{\rm{BH,3}}} = \dfrac{|\vec{\chi}_{\rm{BBH,in}}|}{1+M_{\rm{BH,3}}/M_{\rm{BBH,in}}},
\end{equation}
where $\vec{\chi}$ is the dimensionless component spin of the \ac{BH} and $\hat{L}_{\rm{N}}$ is the unit vector parallel to the system's orbital angular momentum.
For a merger of an equal mass inner non-rotating \ac{BBH}, the remnant has a spin magnitude of $|\vec{\chi}_{\rm{BBH,in}}|\approx 0.68$ \citep{Boyle2008BBHs}. 

\section{Results} \label{sec:results}
\subsection{\ac{BH} mass parameter-space for sequential mergers}
\label{subsec:regions}
\begin{figure}
\centering
\includegraphics[trim=0 2.7cm 0 7cm, clip,width=\textwidth]{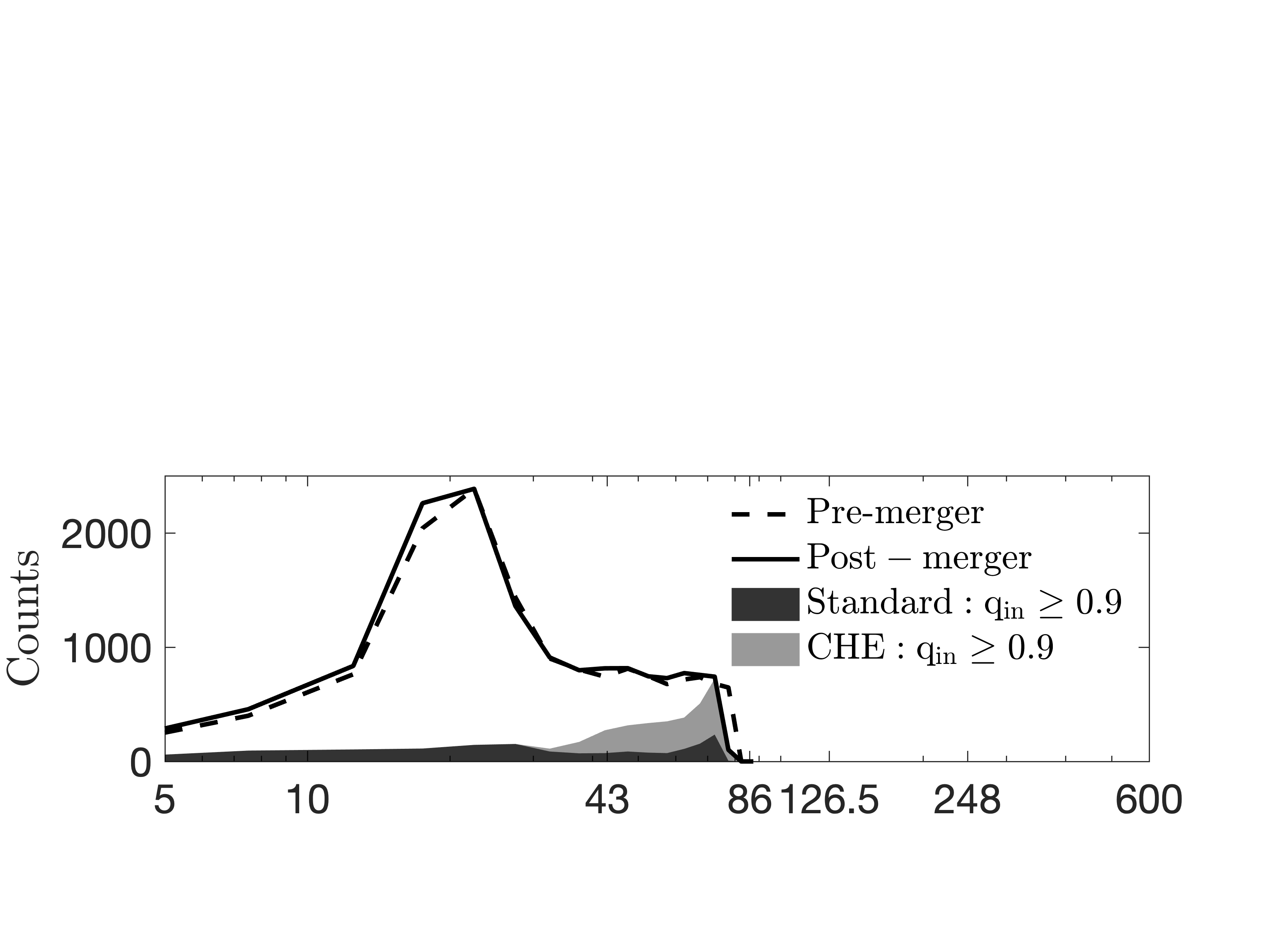}
\includegraphics[trim=0 0 0 0.5cm, clip,width=\textwidth]{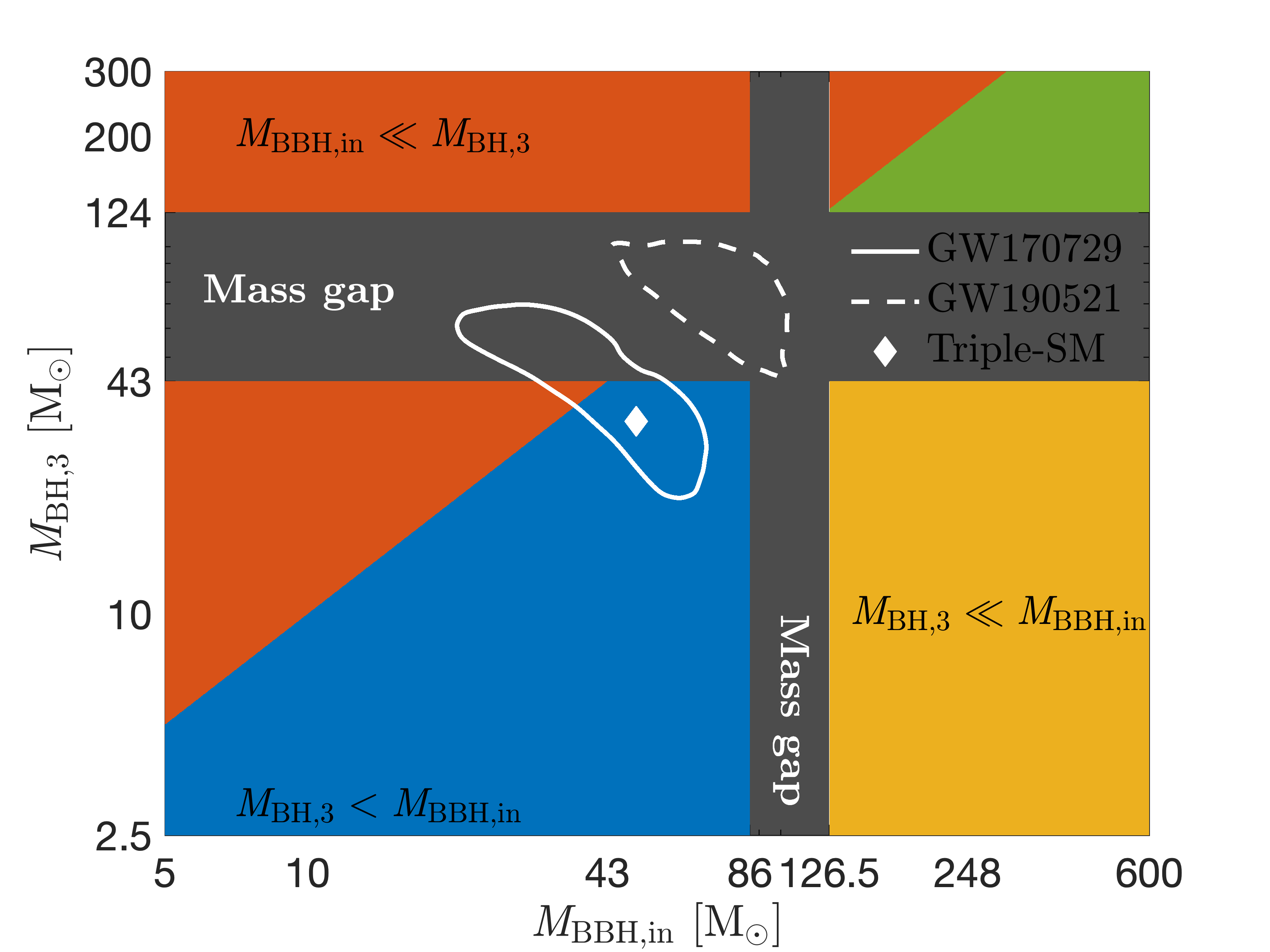}
\caption{
Overview of mass combinations for isolated triples leading to sequential mergers.
Top: 
Mass distribution of merging \acp{BBH} for the inner binary based on isolated binary evolution calculations \citep{Riley2020CHE}.
The solid (dashed) black line is the mass distribution after (before) the inner \ac{BBH} merger.
Shaded regions highlight standard evolving (dark grey) and \ac{CHE} (light grey) systems with $q_{\rm{in}} \ge 0.9$.
Bottom: 
The area colored grey is the approximate mass gap region where \acp{BH} from isolated binary origin are not expected to form due to \acp{PISN} (Section \ref{subsec:regions}).
The coloured regions are described in Section \ref{subsec:regions} and examples are illustrated in Figure \ref{fig:cartoon}.
The diamond is an example of a sequential merger (Triple-SM) as discussed in Section \ref{subsec:event}.
White solid and dashed contours are the 90\% confidence intervals for \event\ \citep{Chatziioannou2019GW170729} and \eventGap\ \citep{GW190521PRL}, respectively.
}
\label{fig:totalMass}
\end{figure}

In Figure \ref{fig:totalMass} we display the possible mass combinations of inner \acp{BBH} and outer \acp{BH} for sequential mergers. 
We denote the (\ac{PISN}) mass gap in grey. 
Assuming single stellar evolution for the tertiary, the mass gap for the outer \ac{BH} is between $\rm \textit{M}_{gap,min} = 43\ M_{\odot} \lesssim \textit{M}_{BH,3} \lesssim \textit{M}_{gap,max} = 124\ M_{\odot}$.
The corresponding mass gap for the inner \ac{BBH} is shifted to $\rm 2\times \textit{M}_{gap,min} = 86\ \rm{M_{\odot}} \lesssim \textit{M}_{BBH,in} \lesssim \textit{M}_{gap,max}+\textit{M}_{BH,min} = 126.5\ M_{\odot}$.

In Figure \ref{fig:totalMass} we classify four regions of interest: red, blue, yellow and green.
The red region comprises $\rm \textit{M}_{BBH,in} < \textit{M}_{BH,3}$, where the outer (more massive) \ac{BH} collapses before the other \acp{BH} form, based on stellar evolution timescales. 
The blue region comprises $\rm \textit{M}_{BH,3} < \textit{M}_{BBH,in}$ with $\rm \textit{M}_{BBH,in} \lesssim 86\ \rm{M_{\odot}}$ and $\rm \textit{M}_{BH,3} \lesssim 43\ \rm{M_{\odot}}$. 
Most \ac{GW} sources detected to date are in the red and blue region below the mass gap \citep{GWTC12019}.
The yellow region comprises $\rm \textit{M}_{BBH,in} \gtrsim 126.5\ M_{\odot}$ with  $\rm \textit{M}_{BBH,in} \gg \textit{M}_{BH,3}$ and $\rm \textit{M}_{BH,3} \lesssim M_{\rm{gap,min}} \approx 43\ M_{\odot}$. 
Finally, the green region comprises $\rm \textit{M}_{BBH,in} > \textit{M}_{BH,3}$ with $\rm \textit{M}_{BBH,in} \gtrsim 126.5\ M_{\odot}$ and $\rm \textit{M}_{BH,3} \gtrsim 124\ M_{\odot}$. 
The rare sequential mergers from the green region will have masses above the mass gap and are not discussed any further.

\subsection{Types of Triples}
\label{subsec:types}
For the evolution of the stars, we consider \textit{standard evolving} and \textit{compact} constituent stars.
Standard evolving stars rotate slowly, develop a composition gradient, and  can expand up to thousands of solar radii during their evolution \citep[e.g.,][]{Podsiadlowski1992}.
This expansion is avoided by stars in metal-free (Pop III) environments
\citep{Marigo2001PopIII}, in certain mass-metallicity regimes \citep{Shenar:2020} or that rotate rapidly such that rotational mixing is induced \citep{Maeder1987}, i.e. \ac{CHE}.
The orbits of \ac{CHE} binaries can therefore remain more compact throughout their evolution to \ac{BBH} formation as compared to binaries with standard evolving stars \citep{Marchant2016CHE,deMinkMandel2016,Riley2020CHE}. 
This is favorable for the sequential merger channel in order for the triple to remain dynamically stable as well as compact enough to lead to two mergers within a Hubble time (Equation\,\ref{eq:triple}~\&~\ref{eq:Peters}). 
We therefore consider four distinct types of stellar triples (illustrated in Figure \ref{fig:cartoon}): 
1) an inner binary with compact stars and an outer standard evolving star,
2) all compact stars,
3) at least one standard evolving star in the inner binary with an outer standard evolving companion, and,
4) at least one standard evolving star in the inner binary with an outer compact star.

\begin{figure}
\centering
\includegraphics[trim=0 2.7cm 0 7cm, clip,width=\textwidth]{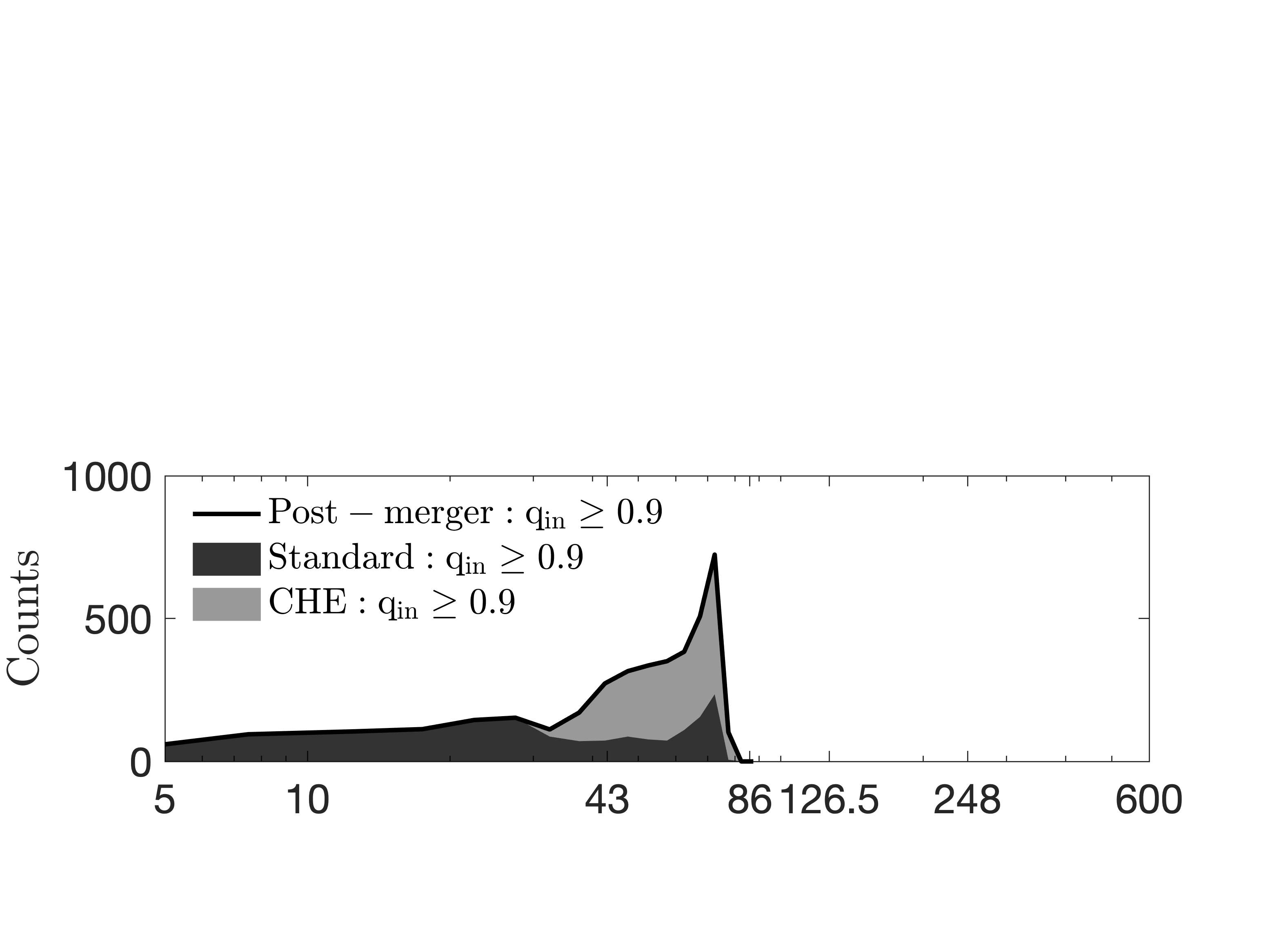}
\includegraphics[trim=0 0 0 0.5cm, clip,width=\textwidth]{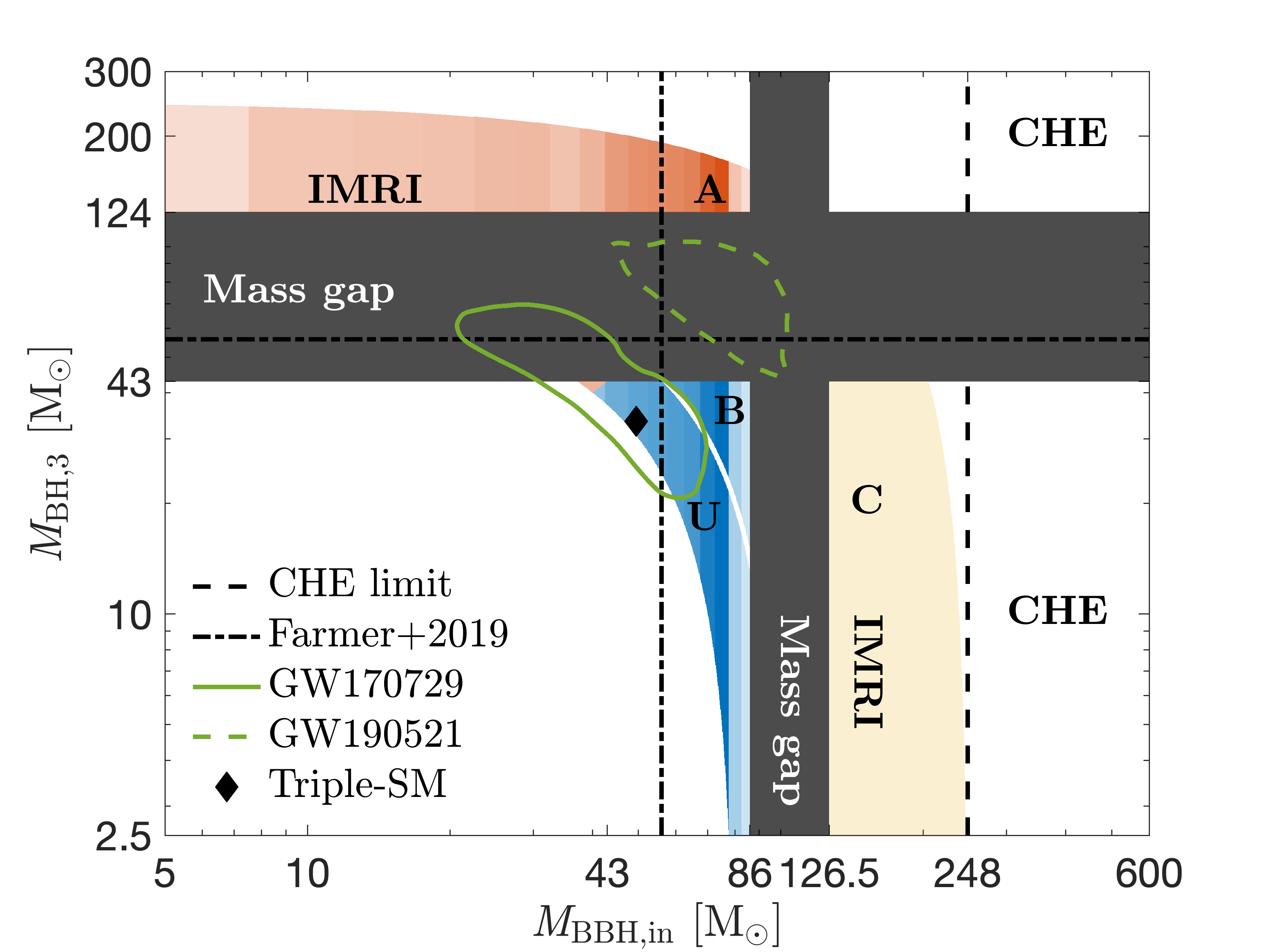}
\caption{
Overview of mass combinations for isolated triples leading to sequential mergers with total mass in the (\ac{PISN}) mass gap.
Similar to Figure \ref{fig:totalMass} but only accounting for sequential mergers that have final total mass within the mass gap.
For the mass of the inner \ac{BBH} the intensity of the color reflects the bin weight from the mass histogram (top panel).
We explore mass gap sequential mergers in the red (A), blue (B) and yellow (C) regions (Section \ref{subsec:PISNgap} for full description).
Blue sub-region B only includes sources well within the mass gap with total mass $(M_{\rm{BBH,in}}+M_{\rm{BH,3}})/\rm{M_{\odot}} > 100$.
To incorporate model uncertainties,  we lower the threshold to between $80 \leq (M_{\rm{BBH,in}}+M_{\rm{BH,3}})/\rm{M_{\odot}} \leq 100$ \citep{Stevenson2019PISN} for region U.
The black dash-dotted line at $M_{\rm BBH,in}= 56\ \rm{M_{\odot}}$ corresponds to the maximum single \ac{BH} mass according to \cite{Farmer2019PISNgap}.
The area encompassed to the right of this limit and the left of the mass gap is not populated by single stellar evolution.
The black dashed line corresponds to $2\times M_{\rm{gap,max}} \approx 248\ \rm{M_{\odot}}$, the minimum combined mass for \ac{CHE} \acp{BBH} above the mass gap \citep{duBuisson2020CHE}.
Intermediate mass-ratio inspirals (IMRIs) can be in the red and yellow regions.
See Figure \ref{fig:totalMass} for further explanations and Figure \ref{fig:cartoon} for an illustration of the evolution of these systems.
}
\label{fig:massGap}
\end{figure}

\begin{figure}
\centering
\includegraphics[trim=0 0 0 0, clip,width=\textwidth]{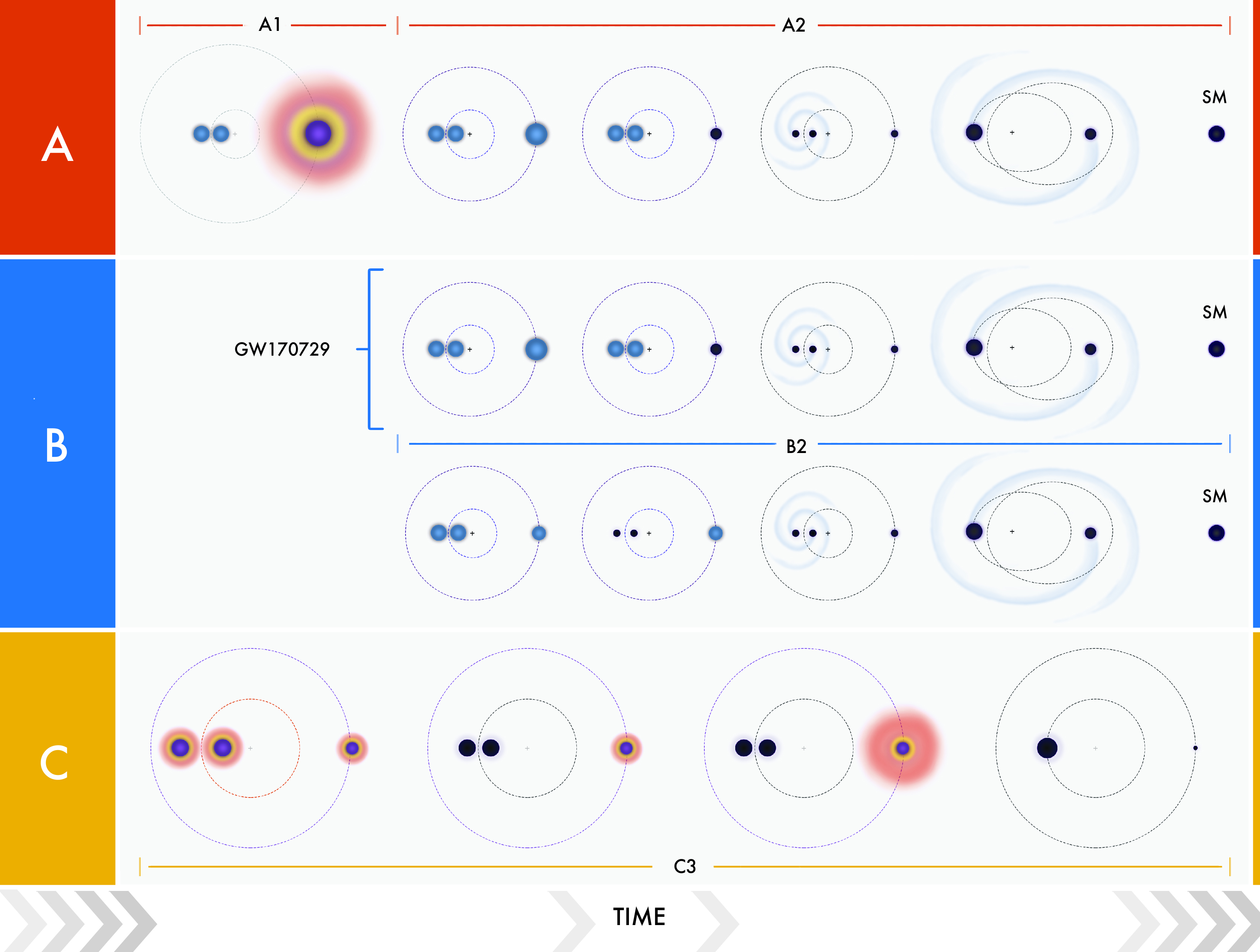}
\caption{
Time evolution of triple stellar systems.
We illustrate compact stars in blue, standard evolving stars with red envelopes, \acp{BH} in black and merging \acp{BBH} with a surrounding swirl.
\acp{BH} formed as sequential mergers are labeled \textit{SM}.
Architectures are described in Sections \ref{subsec:regions} and \ref{subsec:types}, and discussed in Section \ref{subsec:PISNgap}.
Top: 
Red sub-region A, where the tertiary is the most massive star in the system and forms the first \ac{BH} in this triple.
The inner binary needs to be constituted of compact stars and the tertiary can be either standard evolving or a compact star (Triple Types 1 and 2 respectively). 
Middle: 
Blue sub-region B, where all stars have similar masses.
These triples can only lead to sequential mergers if all stars are compact (Triple Type 2). 
We suggest \event\ experienced this evolution.
Bottom: 
Yellow sub-region C, where the tertiary is of significantly lower mass than either of the inner binary stars. 
We find that this configuration (Triple Type 3) does not lead to sequential mergers and is only presented for completion. 
Credit: T. Rebagliato.
}
\label{fig:cartoon}
\end{figure}

\begin{figure}
\includegraphics[width=0.5\textwidth]{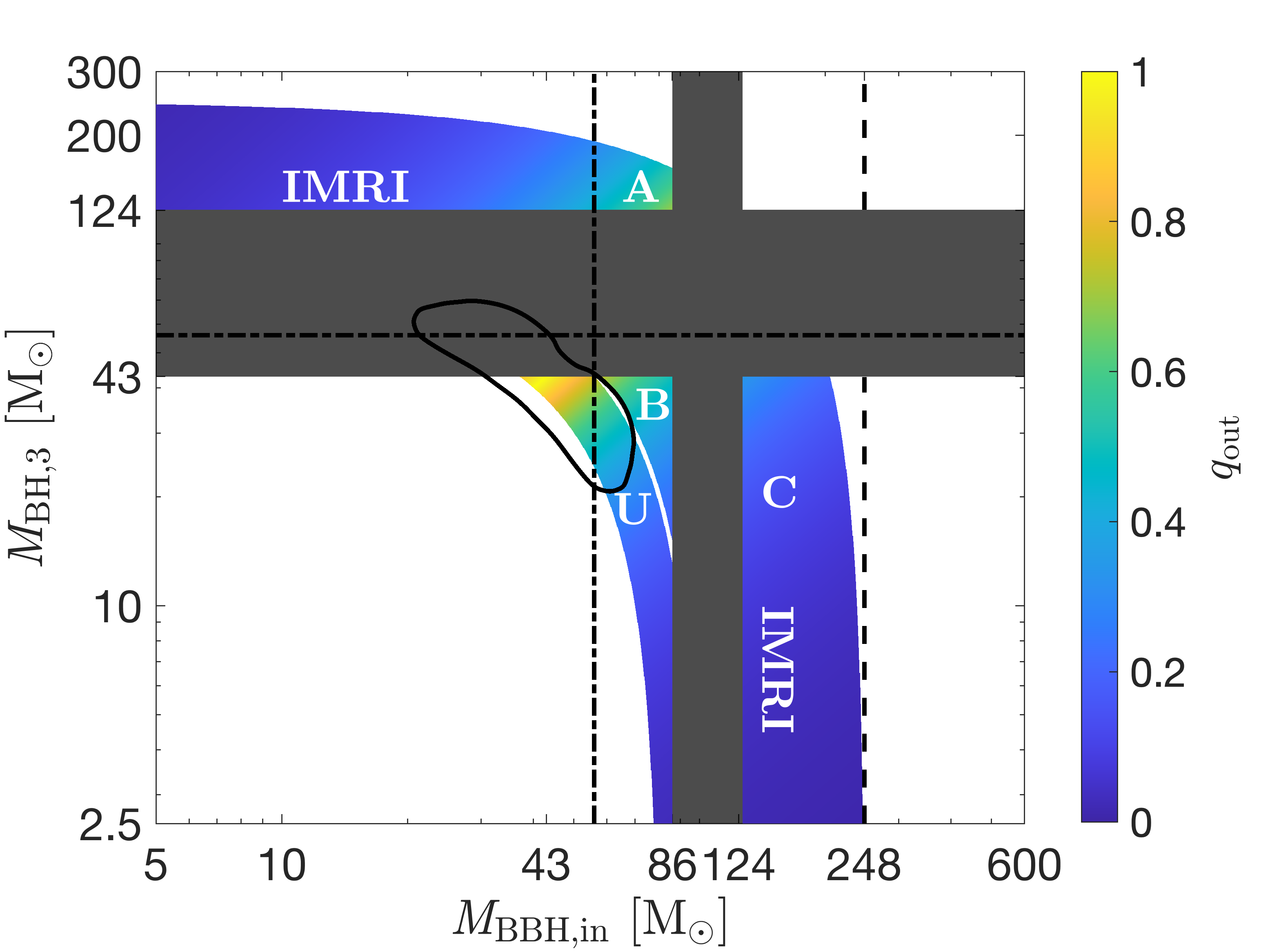}
\includegraphics[width=0.5\textwidth]{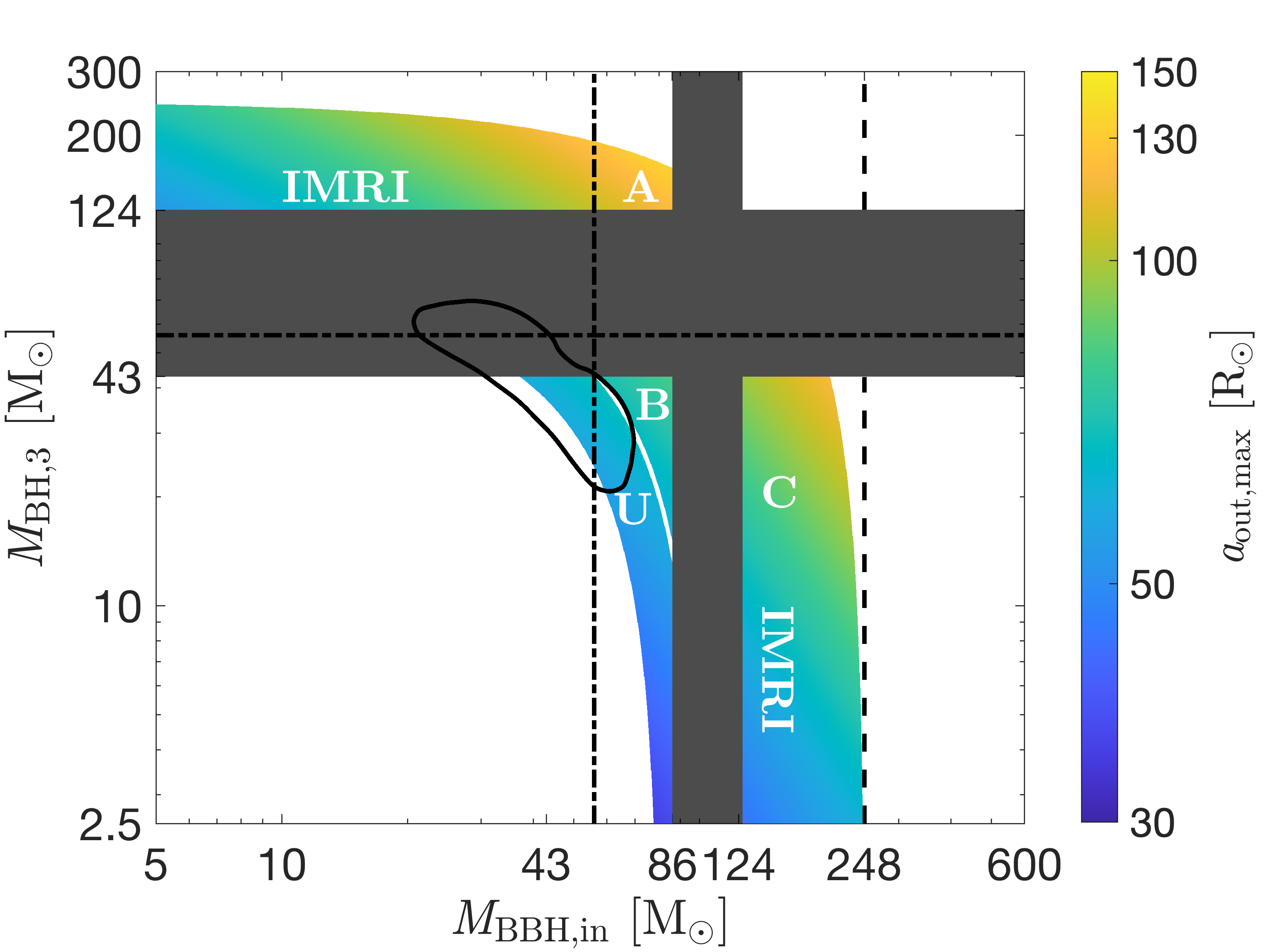}
\includegraphics[width=0.5\textwidth]{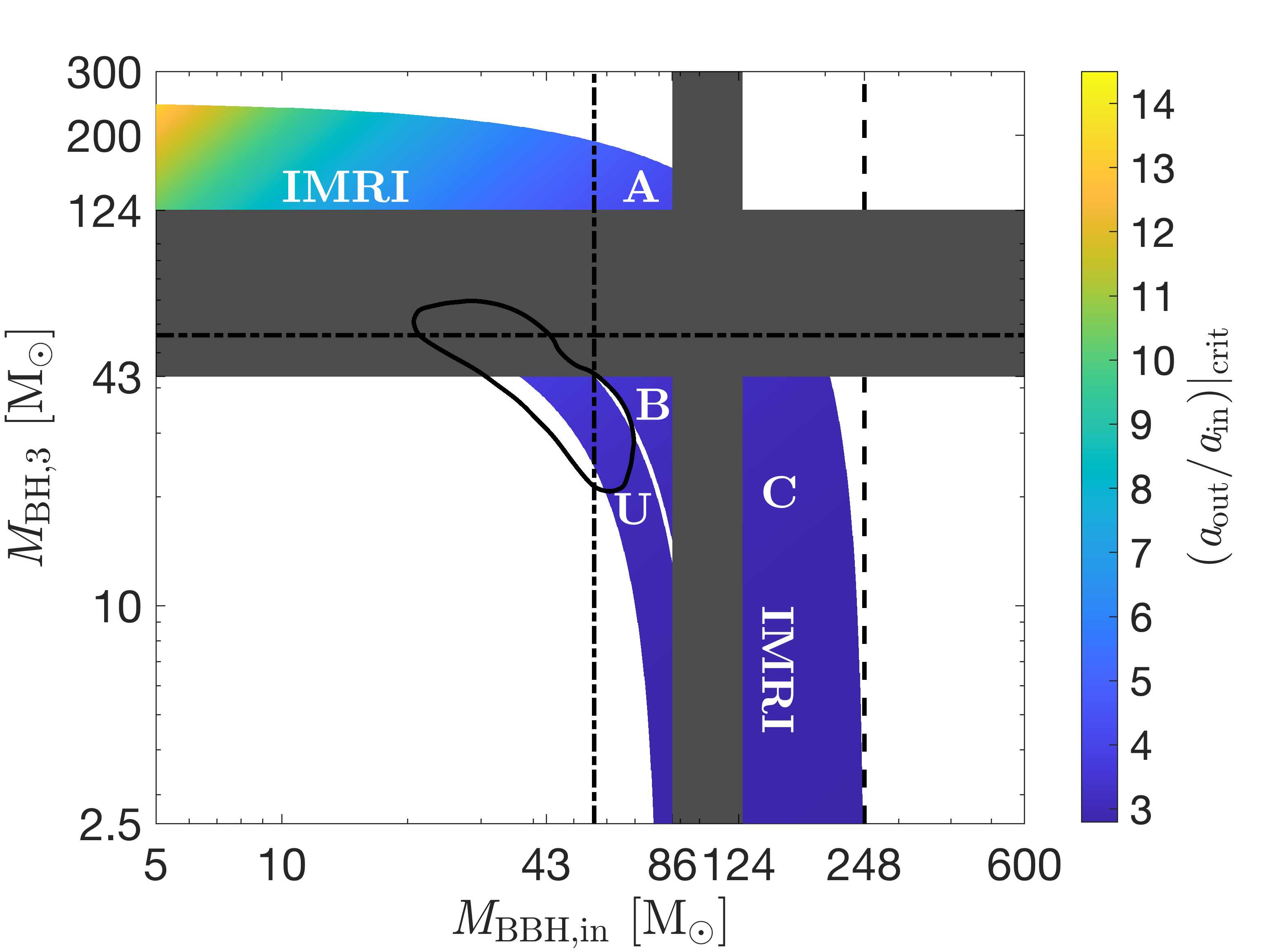}
\includegraphics[width=0.5\textwidth]{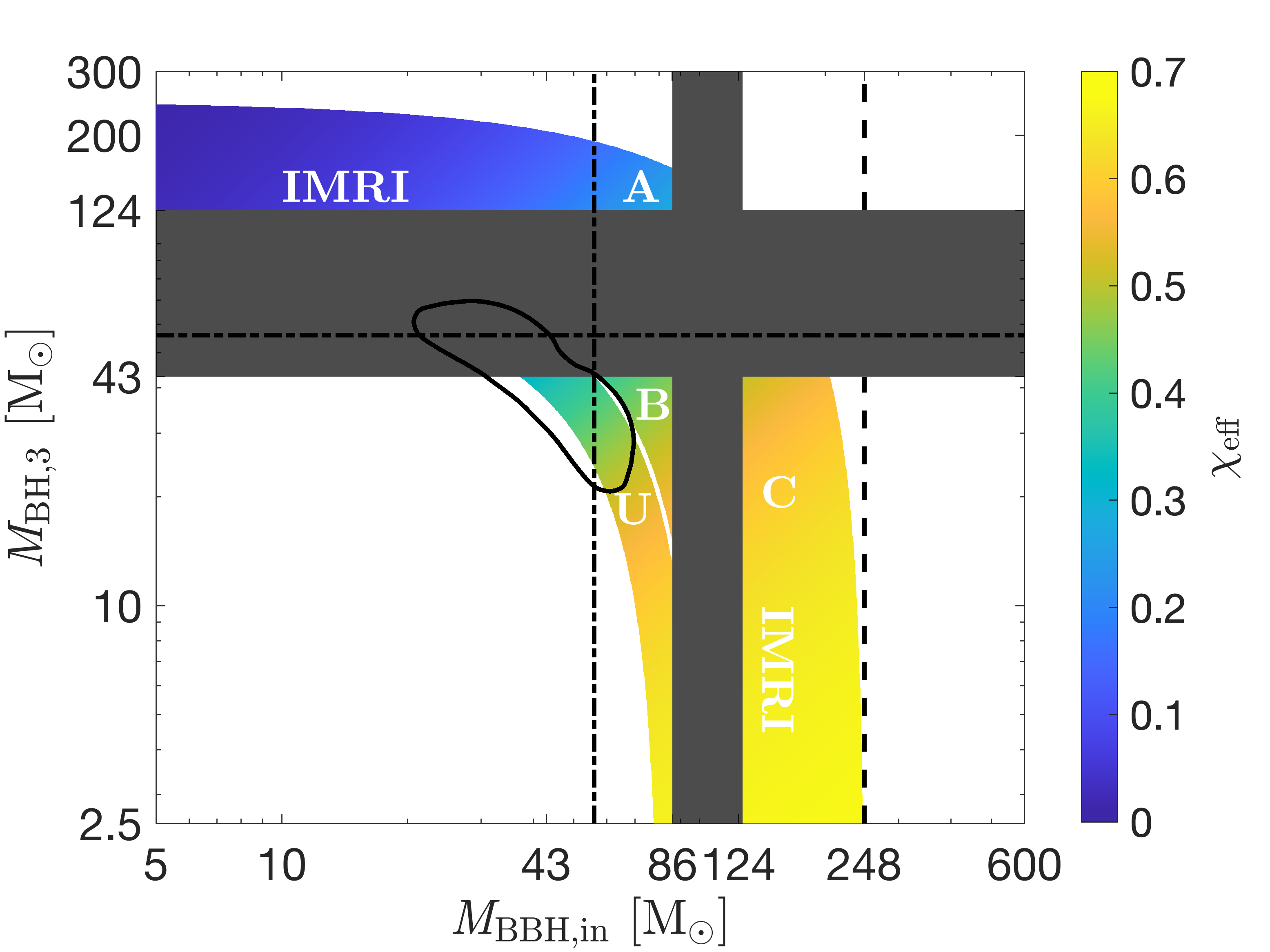}
\caption{
Overview of mass combinations for isolated triples leading to sequential mergers with total mass in the (\ac{PISN}) mass gap (see caption in Figure \ref{fig:totalMass} and \ref{fig:massGap} for additional details).
The color in each panel denotes a different physical property.
90\% confidence intervals for \event\ \citep{Chatziioannou2019GW170729} shown as a solid black line.
Top left: outer mass ratio ($q_{\rm{out}}$) for the sequential merger. 
Top right: maximum separation in which a circular binary with component masses $M_{\rm{BBH,in}}$ and $M_{\rm{BH,3}}$ can merger within the age of the Universe due to \ac{GW} emission.
Bottom left: minimum ratio of outer and inner orbital separation for dynamical stability assuming a circular coplanar prograde orbit.
Bottom right: effective dimensionless spin ($\chi_{\rm{eff}}$) of the sequential merger, assuming coplanarity in the inner and outer orbit. Note that $\chi_{\rm{eff}} > 0$ and $\chi_{\rm{eff}}\approx 0.4$ for \ac{LVC}-like sources (around sub-region B).
}
\label{fig:totalMassProperties}
\end{figure}

\begin{table*}
\caption{
Summary of expected demographics of different types of  sequential mergers. 
This Table summarizes the results from Section \ref{sec:results} and all Figures.
For each quantity of interest we present the minimum and maximum values in between square brackets. 
We have subjectively provided a ranking for most likely (1) to less likely (5) scenario to form a sequential merger, based on the inner \ac{BBH} mass distribution, the condition for triple stability and the sequential merger time.
Masses and separations are in solar units. 
}
\centering
\begin{tabular}{@{}cccccc@{}}
\hline\hline
Region & A-IMRI & A & U & B & C-IMRI \\
\hline\hline
Rank & 4 & 2 & 1 & 3 & 5 \\
Region & Red & Red & Blue & Blue & Yellow \\
Triple type & 1 & 1 \& 2 & 1 \& 2 & 1 \& 2 & 3 \\
Mass gap $M_{\rm{BBH,in}}$ & No & Yes & Uncertain & Yes & No \\
$M_{\rm{BH,3}}$ & [124,243] & [124,205] & [2.5,43] & [14,43] & [2.5,43] \\
$M_{\rm{BBH,in}}$ & [5,43] & [43,86] & [37,86] & [57,86] & [126.5,245.5] \\
$M_{\rm{BH,3}}+M_{\rm{BBH,in}}$ & [129,248] & [167,248] & [80,100] & [100,129] & [129,248] \\
$q_{\rm{out}}$ & [0,0.1] & [0.1,0.69] & [0,1] & [0.16,0.74] & [0,0.1] \\
$a_{\rm{out}}$ & [52,120] & [97,135] & [35,69] & [57,82] & [45,120] \\
$\chi_{\rm{eff}}$ & [0,0.1] & [0.1,0.27] & [0.4,0.68] & [0.38,0.58] & [0.5,0.68] \\
$(a_{\rm{out}}/a_{\rm{in}})|_{\rm{crit}}$ & [4.8,13.3] & [4.0,5.7] & [2.8,3.8] & [3.0,3.3] & [2.8,3.1]
\end{tabular}
\label{tab:summary}
\end{table*}

\subsection{Populating the (\ac{PISN}) mass gap}
\label{subsec:PISNgap}
From the full \ac{BH} mass range (Figure \ref{fig:totalMass}) we extract sequential mergers with total mass within the mass gap, classify them in sub-regions (Figure \ref{fig:massGap}), discuss their evolutionary pathways (Figure \ref{fig:cartoon}), and present their demographics  (Figure \ref{fig:totalMassProperties} and Table \ref{tab:summary}).

\subsubsection{Red sub-region A}
\label{subsubsec:red}
The predicted properties of systems in red sub-region A are $\rm \textit{M}_{BBH,in} \ll M_{\rm{gap,max}} \leq \textit{M}_{BH,3}$ and $0.1 \lesssim \chi_{\rm{eff}} \lesssim 0.27$.
The evolution of an example system is illustrated in the top panel of Figure \ref{fig:cartoon}.
Consider a \ac{CHE} inner binary (Triple Type 1 or 2) with $M_{\rm{1}}\approx M_{\rm{2}} \approx 40\ \rm{M_{\odot}}$, $R_{\rm{1}}\approx R_{\rm{2}} \approx 6\ \rm{R_{\odot}}$ and $a_{\rm{in}} \gtrsim 18\ \rm{R_{\odot}}$.
All stars in this triple will experience complete fallback, which effectively leave the inner and outer orbits unchanged (Section \ref{sec:method}).
Assuming the outer star collapses to a \ac{BH} with mass $M_{\rm{BH,3}} \approx 140\ \rm{M_{\odot}}$, then $a_{\rm{out}} \gtrsim (a_{\rm{out}}/a_{\rm{in}})|_{\rm{crit}}\times a_{\rm{in}} \approx 4.2 \times (18\ \rm{R_{\odot}}) \approx 76\ \rm{R_{\odot}}$ in order for the triple to be stable (Equation\,\ref{eq:triple} and
Figure\,\ref{fig:totalMassProperties}). 
If the separation after the inner \ac{BBH} merges is less than $a_{\rm{out,max}}\lesssim 122\ \rm{R_{\odot}}$, then the sequential merger can occur within a Hubble time (Equation \ref{eq:Peters}). 
The critical ratio of $a_{\rm out}/a_{\rm in}$ to maintain stability and the maximum orbital separation to achieve a merger within a Hubble time depend on the mass combinations of the triple, and vary within a factor of a few for the combinations of interest here (Figure\,\ref{fig:totalMassProperties}).

If the outer star was initially a standard evolving star (Triple Types 1 and 3) it might initiate a mass transfer phase onto the inner binary early in the evolution of the system (step A1 in Figure\,\ref{fig:cartoon}).
This could occur if the radius of a standard evolving star exceeds the Roche lobe radius, i.e. $R_3  \gtrsim 0.43 \times~a_{\rm{in}}$ for $q_{\rm{out}} \approx 140/80$.
Assuming the stellar radii approach $\sim 100-1000\ R_{\odot}$ at maximum \citep{Podsiadlowski1992,Riley2020CHE}, tertiary driven mass transfer occurs for outer orbits up to several $\sim 100-1000\ R_{\odot}$.

During this mass transfer phase, where the tertiary donor is significantly heavier than the inner binary, the outer orbit likely shrinks (based on angular momentum considerations) and the hydrogen envelope is stripped off the outer star. 
If the inner binary avoids a merger during this stage \citep[see, e.g.,][]{LeighToonen2020Triples}, at the end of the mass transfer phase, the tertiary is a stripped helium star reminiscent of the compact outer star considered in Triple Type 2.

In this section so far we have considered inner binaries comprising of \ac{CHE} stars which remain close to each other throughout their evolution to \acp{BBH}.
However, for standard evolving inner binaries (Triple Types 3 and 4), the orbital separation can change drastically between the zero-age main sequence and \ac{BBH} formation due to mass and angular momentum exchanges during mass transfer episodes.
From our synthetic population, the orbits of standard evolving binaries can remain as small as $a_{\rm{in}} \approx 70\ \rm{R_{\odot}}$ throughout their evolution, but for most systems their orbits expand to hundreds or thousands of solar radii \citep[see, e.g.][]{LeighToonen2020Triples}.
Assuming the optimistic case of a standard evolving inner binary with $a_{\rm{in}} \approx 70\ \rm{R_{\odot}}$ maximally, the outer separation must be $a_{\rm{out}}\ge 2.8 \times (70\ \rm{R_{\odot}}) \approx  196\ \rm{R_{\odot}}$ in order for the triple to be stable throughout its full evolution. 
Such an orbit does not merge by \ac{GW} emission alone in a Hubble time (Figure \ref{fig:totalMassProperties}). 
We conclude that in red sub-region A only triples with inner compact binaries can lead to sequential mergers within a Hubble time.

\subsubsection{Blue sub-region B}
The predicted properties of systems in blue sub-region B are  $\rm \textit{M}_{BBH,in}/2\approx \textit{M}_{BH,out} \leq M_{\rm{gap,min}}$ and $0.38 \lesssim \chi_{\rm{eff}} \lesssim 0.58$.
The evolution of an example system is illustrated in the middle panel of Figure \ref{fig:cartoon}.
Furthermore, the synthetic population suggests $\rm \textit{M}_{BH,1}\approx \textit{M}_{BH,2}$ in this region; with similar \ac{BH} masses the evolutionary timescales for all component stars are similar as well. 
Consider a triple with $M_{\rm{1}}\approx M_{\rm{2}} \approx M_{\rm{3}} \approx 40\ \rm{M_{\odot}}$. 
We again assume a \ac{CHE} inner binary (Triple Type 1 or 2) with $R_{\rm{1}}\approx R_{\rm{2}} \approx 6\ \rm{R_{\odot}}$ and $a_{\rm{in}} \gtrsim 18\ \rm{R_{\odot}}$. This triple is dynamically stable if $a_{\rm{out}} \gtrsim 3.3 \times (18\ \rm{R_{\odot}}) \approx 60\ \rm{R_{\odot}}$.
The outer separation must be $a_{\rm{max}}\lesssim 76\ \rm{R_{\odot}}$ for the sequential merger to occur within a Hubble time due to GW emission.
Hence the small possible range for outer separations between $60 \lesssim a_{\rm{out}}/\rm{R_{\odot}} \lesssim 76$ constraints the feasibility and frequency of this sub-channel.  
Pop III \ac{BBH} progenitors with initial masses $M_1 \gtrsim 40\ \rm{M_{\odot}}$ and $q \approx 1$ have initial separations $a_{\rm{in}} \gtrsim 20\ \rm{R_{\odot}}$ \citep{Inayoshi2017PopIIIBBHs}, which lead to slightly more stringent constraints than for \ac{CHE} binaries.  

If the tertiary was a standard evolving star (Triple Types 1 and 3), it could initiate a mass transfer episode onto the inner binary. Due to the mass ratio, we expect this mass transfer to proceed in a stable manner. Furthermore, we expect the outer orbit to widen typically due to angular momentum evolution, which makes the sequential merger less likely to occur within a Hubble time \citep[see, e.g.,][for an overview in the context of BBH progenitors]{belczynski2002comprehensive}. 
In summary, in the blue sub-region B only triple compact binaries in a fine tuned configuration can lead to sequential mergers within a Hubble time.

\subsubsection{Yellow sub-region C}
The masses of systems in yellow sub-region C are  $\rm \textit{M}_{BH,3} \ll M_{\rm{gap,max}} \le \textit{M}_{BBH,in}$.
The evolution of an example system is illustrated in the bottom panel of Figure \ref{fig:cartoon}.
Even though our synthetic binary population does not predict any inner \acp{BBH} in this region (Appendix), \acp{BBH} above the mass gap have been suggested for both compact \citep{Marchant2016CHE,duBuisson2020CHE} and standard evolving \citep{Mangiagli2019gap} binaries.
Regarding the former, as inner binaries experiencing \ac{CHE} (Triple Types 1 or 2) are expected to have mass ratios $q_{\rm{in}}\approx 1$ and $M_{\rm{BBH,in}} \gtrsim 2\times M_{\rm{gap,max}} \approx 248\ \rm{M_{\odot}}$, these systems can only bring about sequential mergers with a total mass above the mass gap.
Regarding the latter, the case of hypothetical standard evolving binaries is also not promising for sequential mergers.
If mass transfer would take place from the lower-mass tertiary companion to the heavier-mass inner binary, it would likely be stable and widen the outer orbit. 
This would increase the \ac{GW} inspiral time for the outer orbit. 
Therefore, we do not consider sequential mergers from sub-region C to be common.

\subsection{\event\ as a sequential merger}
\label{subsec:event}
\event, with a reported chirp mass $\mathcal{M}=(M_1M_2)^{3/5}/(M_1+M_2)^{1/5}= 35.4_{-4.8}^{+6.5}\ \rm{M_{\odot}}$, post-merger remnant \ac{BH} mass of $79.5_{-10.2}^{+14.7}\ \rm{M_{\odot}}$, $\chi_{\rm{eff}}=0.37_{-25}^{+21}$ and redshift $z=0.49^{+0.19}_{-0.21}$ \citep{GWTC12019}, can be marginally considered in the mass gap (Figures \ref{fig:totalMass}, \ref{fig:massGap}, and \ref{fig:totalMassProperties}).
While \cite{Rodriguez2019Gap} associates \event\ with cluster origin, \cite{KimballTalbot2020} claims a standard stellar evolution origin. 
Here we propose that the heavier \ac{BH} of \event, with an inferred mass of $50.2_{-10.2}^{+16.2}\ \rm{M_{\odot}}$, is the remnant of an inner \ac{BBH} from a sequential merger.

The scenario we suggest is the following.
We consider a \ac{CHE} system (Triple Type 3) at metallicity $Z=10^{-4}$ with an inner over-contact binary of $M_1=M_2=29\ \rm{M_{\odot}}$, $R_1=R_2\approx 4.8\ \rm{R_{\odot}}$, $a_{\rm{in}}\approx 10.5\ \rm{R_{\odot}}$ and a tertiary companion with $M_3=33.5\ \rm{M_{\odot}}$,  $a_{\rm{out}}=43.5\ \rm{R_{\odot}}$ and $e_{\rm{out}}=0$. 
This satisfies the stability condition for the triple as $a_{\rm{out}} > 3.4 \times (10.5\ \rm{R_{\odot}}) \approx 35\ \rm{R_{\odot}}$ (Figure \ref{fig:totalMassProperties}).
At that time, the outer orbit has a relative speed of $\sqrt{G (M_1+M_2+M_3)/a_{\rm{out}}} \approx 600\ \rm km\ s^{-1}$.
All stars in this triple will experience complete fallback and negligible stellar winds, which effectively leave the inner and outer orbits unchanged.
The only mass change during \ac{BH} formation comes from the relation between baryonic and gravitational mass \citep{fryer2012compact}.
The first star to collapse to a $\approx 30.2\ \rm{M_{\odot}}$ \ac{BH} is the tertiary at around 6 Myr. 
Subsequently the inner binary evolves into a $\approx 26+26\ \rm{M_{\odot}}$ \ac{BBH} at around 7 Myr. 
Afterwards evolution is purely driven by \ac{GW} radiation. 
It takes the inner binary $\sim$50 Myr to merge and form a single \ac{BH} with $M_{\rm{BBH,in}}\approx (1-f_{\rm{rad}})(M_{\rm{BH,1}}+M_{\rm{BH,2}}) \approx 49.5\ \rm{M_{\odot}}$, spin $|\vec{\chi}_{\rm{BBH,in}}|=0.68$ \citep{Boyle2008BBHs}, separation $a_{\rm{out}}\approx 45\ \rm{R_{\odot}}$ (Equation \ref{eq:BlaauwSeparation}) and $e_{\rm{out}}\approx 0$ (Equation \ref{eq:BlaauwEccentricity}). 
After $\sim 5.1$ Gyr (Equation \ref{eq:Peters}) the second merger takes place. This sequential merger has a chirp mass of $\approx 33.5\ \rm{M_{\odot}}$, $\chi_{\rm{eff}}\approx 0.4$ (Figure \ref{fig:totalMassProperties}) and $z\approx 0.49$ (Appendix for redshift calculation and details on the evolution) and is shown in Figures \ref{fig:totalMass} and \ref{fig:massGap} as a diamond (Triple-SM).

\subsection{\eventGap: a \ac{BBH} in the mass gap}
\label{subsec:eventGap}
The inferred total mass of \eventGap\ is within the mass gap with \ac{BH} component masses of $85_{-14}^{+21}\ \rm{M_{\odot}}$ and $66_{-18}^{+17}\ \rm{M_{\odot}}$ \citep{GW190521PRL,GW190521ApJL}.
The heavier \ac{BH} of \eventGap\ has a mass greater than $60\ \rm{M_{\odot}}$ \citep{Fishbach2020GW190521} and therefore can not be easily associated to isolated binaries (Section \ref{subsec:uncertaintiesPISNmassgap}).
While uncertainties in the inferred component masses make it tempting to associate \eventGap\ with a triple origin the inferred spins suggest otherwise.
The high individual spins ($|\vec{\chi}_{1,2}|>0.69$) and low effective spin ($\chi_{\rm{eff}}\approx 0$) imply that \ac{BH} spins are not aligned with the orbit (Equation \ref{eq:spins}).
We conclude that \eventGap\ is inconsistent with our predictions of sequential mergers.

\section{Discussion} 
\label{sec:disc}
\subsection{Main caveats to our method}
\label{subsec:caveats}
The main caveats of the scheme presented in this {\it Letter} are the natal properties of \acp{BH} and the assumed orbital distributions (Section\,\ref{subsec:stellarTriples}).
The remnant masses, birth spins and natal kicks of \acp{BH} remain an open question in astrophysics.
The assumption of low (Pop II and III) metallicity and complete fallback make mass loss and natal kicks negligible \citep{fryer2012compact,Muller2016}.
This is favorable for sequential mergers because any potential widening of the inner orbit can trigger a dynamical instability of the triple and increase the \ac{GW} inspiral time. 
However, some models of \ac{CHE} binaries predict Wolf-Rayet mass loss is non-negligible \citep[e.g.,][]{Marchant2016CHE}.
Complete fallback is less likely for \ac{BH} progenitors with carbon-oxygen masses $\lesssim 11\ \rm M_{\odot}$ \citep{fryer2012compact}, some of which are associated to \ac{GW} sources \citep[see, e.g., GW170608 from][]{GWTC12019}.

Natal kicks could tilt the orbital spins, change eccentricities and induce inclinations, modifying the evolution from the moment the first \ac{BH} is born.
In some cases, the direction and magnitude of the kick could decrease the inner \ac{BBH} merger timescale; in others, it could ionize the triple before the sequential merger (other effects discussed in Section \ref{subsec:stellarTriples}).
The extreme case of natal kicks with velocities $\gtrsim 100\ \rm km\ s^{-1}$ could even disrupt the system at \ac{BH} formation.

The spin transition from star to \ac{BH} is also quite uncertain.
Recent studies suggest that angular momentum transport \citep{FullerLinhao2019} and accretion feedback \citep{BattaRamirezRuiz2019} disfavor maximally spinning \acp{BH}.
While most \acp{BH} are probably born with low spins, \ac{CHE} can lead to moderate and even high spins \citep{FullerLinhao2019}.
Spin alignment with the orbital angular momentum vector (e.g., \ac{CHE} binaries) prevents a \ac{GW} recoil kick even for maximally spinning \acp{BH}.
Moderate and high misaligned spins can lead to \ac{GW} recoil kicks of tens or hundreds of $\rm km\ s^{-1}$ \citep{Lousto2010}.
These kicks would not necessarily disrupt our systems (e.g., Section \ref{subsec:event}).

We assume no rotation at birth in order to establish lower limits on effective spins, but remain agnostic on what should a realistic natal \ac{BH} kick and spin distribution be.

\subsection{Stellar triples: birth distributions and orbital evolution}
\label{subsec:stellarTriples}
The expected rate of sequential mergers depends on the abundance of massive triples and their architectures.
Even though these are unconstrained at low metallicities, observations in the Local Group indicate that for O-type stars between $16\leq  M/\rm{M_{\odot}} \leq 40$, the triple multiplicity fraction is $\approx 0.35$ \citep{MoeDiStefano2017}. 
Multiplicity in massive stars affects the evolutionary outcomes of field systems, and isolated binary, triple and quadruple evolution jointly contribute to, e.g., the double compact object merger rate.

Close observed systems ($P_{\rm{orb}}\lesssim 1\ \rm{AU}$) are preferably circular \citep{Sana2012,MoeDiStefano2017}.
Observations of tight ($\lesssim 50\ \rm{AU}$) low and intermediate mass ($\lesssim 4\ \rm{M_{\odot}}$) triples  show a strong tendency towards orbital alignment ($\lesssim 36^{\circ}$) and are therefore likely to avoid Lidoz-Kozai cycles \citep{Tokovinin2017Triple}.
However, the orbital distributions of massive stellar triples are yet unknown.
Diverging from our simplified assumption of circular coplanar prograde orbits might abruptly modify the evolution of the \ac{BH} triple due to three-body dynamics. 
This would induce high(er) eccentricities \citep{Naoz2016} in the inner binary that may lead to exchanges of mass, mergers \citep{IbenTutukov1999,MoeDiStefano2017} and disruptions \citep[e.g.,][]{HePetrovich2018}.
Additionally, mass transfer initiated from the tertiary could have the gaseous envelope change the characteristics of the inner binaries with respect to our adopted synthetic population \citep{deVries2014}.
It may even provoke a merger of the inner binary and provide an electromagnetic counterpart \citep{LeighToonen2020Triples}. 
In general, mergers between the two stars of a triple can lead to different systems than the ones expected from pristine binaries \citep[e.g.,][]{Podsiadlowski1992,VignaGomez2019}.
It is therefore likely that triples play a non-negligible role in massive stellar evolution. 

Eccentricity of the outer orbit is an additional caveat. 
While the \ac{GW} coalescence timescale is significantly shorter for eccentric binaries \citep{peters1964gravitational}, the minimum ratio of $a_{\rm out}/a_{\rm in}$ to guarantee dynamical stability is larger for more eccentric systems \citep{MardlingAarseth2001}.
Both of these directly affect the number of sequential mergers.

Future observations of massive stars will shed light on the orbital birth distributions of triples and will help constrain the validity of our assumptions, the rate of sequential mergers and constrains on their observables.

\subsection{Uncertainties in the (\ac{PISN}) mass gap}
\label{subsec:uncertaintiesPISNmassgap}
The exact location and existence of the mass gap is an open question\citep[see, e.g.,][and references therein]{Belczynski2020}. 
Any model uncertainties on the limits of the mass gap directly propagate to the predicted \ac{BBH} mass distribution of sequential mergers. 
Metal-free (Pop III) stars, which can be compact and are believed to lead to more massive \ac{BH} progenitors, have been suggested to shift the edge of the mass gap \citep[e.g.,][]{Farrell2020,Tanikawa2020}.
Helium stellar models predict that the location of the lower edge of the mass gap is robust against variations in (Pop II) metallicity, treatment of rotational mixing and wind mass loss but sensitive to nuclear reaction rates giving $40 \lesssim M_{\rm{gap,min}}/\rm{M_{\odot}} \lesssim 56$ \citep{Farmer2019PISNgap}.
Variation on the $^{12}\rm{C} (\alpha,\gamma)^{16}\rm{O}$ rate could shift the limit to $M_{\rm{gap,min}}/\rm{M_{\odot}} \lesssim 56-90$ \citep{Farmer2020PISNgap} and even make the mass-gap disappear \citep{Costa2020}.
Different supernova prescriptions lead to an uncertainty between $40 \lesssim M_{\rm{gap,min}}/\rm{M_{\odot}} \lesssim 50$ \citep{Stevenson2019PISN}. 
Mass transfer in binaries does not significantly affect the location of the mass gap \citep{vanSon2020massGap}. 
\ac{GW} sources are an exciting prospect to study the mass gap, with the caveat that they must be first segregated into field or cluster origin.

\subsection{Multiple \ac{BH} mergers}
Mass gap mergers are usually associated with multiple \ac{BH} mergers in clusters \citep[e.g. GW190521 as discussed in][and references therein]{GW190521ApJL}.
Such mergers are expected to occur readily when central densities are high \citep{SamsingHotokezaka2020}. 
Multiple \ac{BH} mergers lead to effective spins between $-0.5 \lesssim \chi_{\rm{eff}} \lesssim 0.5$ \citep{Rodriguez2019Gap}.
The rates of multiple \ac{BH} mergers decrease significantly for spinning \acp{BH}, as their merger experiences a recoil kick of magnitude comparable or larger to the escape speed of the cluster \citep{Rodriguez2019Gap}. 
On the other hand, a massive stellar field triple is less likely to be disrupted during the inner \ac{BBH} merger (Section \ref{sec:method}).
A pile-up of \ac{GW} sources in the mass gap with $\chi_{\rm{eff}} \gg 0$ could be an indicative of isolated triple origin. Additionally, systems in which both \ac{BH} have masses above the mass gap and $\chi_{\rm{eff}}<0$ should be seriously considered of cluster origin.

Multiple stellar systems have also been associated to \ac{BBH} mergers.
Standard evolving hierarchical triples of field or cluster origin can lead to an inner \ac{BBH} merger by perturbations of the third companion \citep{Fragione2020MNRAS,Martinez2020Triples}.
Standard evolving quadruple systems have also been suggested as progenitors of \ac{BBH} mergers.
Some of them can be composed of two \ac{BBH} binaries, where recoil kick of the merger remnant of one \ac{BBH} triggers the interaction with the other \ac{BBH}, likely exchanging a component and eventually leading to a \ac{BBH} merger \citep{Fragione2020ApJ}.
Others can be hierarchical quadruples where the Lidov-Kozai effect assists the sequential merger of the two inner binaries \citep{Safarzadeh2020ApJ}.

Alternatively, some \acp{BBH} orbiting around galactic nuclei can experience Lidov-Kozai perturbations from the central super-massive black hole, increasing the \ac{BBH} merger rate \citep{Hoang2018ApJ}.

\subsection{Intermediate mass-ratio inspirals}
\label{subsec:IMRIs}
Sequential mergers with $M_{\rm{BBH,in}} > 100\ \rm{M_{\odot}}$ and $M_{\rm{BH,3}} < 43\ \rm{M_{\odot}}$, which correspond to the red and yellow region above the mass gap in Figures \ref{fig:totalMass} and \ref{fig:massGap}, are \acp{IMRI} and can be degenerate with those of binary origin (Appendix). 
\acp{BH} with $M_{\rm{BH}} \gtrsim 100\ \rm{M_{\odot}}$ are usually associated with clusters \cite[see][for a review]{MillerColbert2004IMBHs} but can also be formed by single massive stars \citep{Heger2003}. 
\acp{IMRI} have not yet been detected, but the merger product of \eventGap is the first detected intermediate-mass \ac{BH} \citep{GW190521PRL}. 

For our assumptions we expect $\chi_{\rm{eff}}=0$ for \acp{IMRI} of binary origin, and therefore $\chi_{\rm{eff}} \gg 0$ is a strong indicative of sequential merger origin.
\acp{IMRI} in the yellow region have $0.5 \lesssim \chi_{\rm{eff}} \lesssim 0.68$  but \acp{IMRI} in the red region have $0 \lesssim \chi_{\rm{eff}} \lesssim 0.1$ 
(Figures \ref{fig:totalMass}, \ref{fig:massGap}, \ref{fig:totalMassProperties} and Table \ref{tab:summary}).

\subsection{Rate of sequential mergers}
At redshift $z=0$, \cite{Riley2020CHE} predicts a \ac{BBH} merger rate of $50\ \rm Gpc^{-3}\ yr^{-1}$, which we use to estimate a sequential merger rate $\mathcal{R} < 3\ \rm Gpc^{-3}\ yr^{-1}$ from field triples (Appendix), similar to the estimated rate from hierarchical triples in clusters \citep{Martinez2020Triples} and in the field  \citep{Antonini2017ApJ}.

For sequential mergers, the inner compact \ac{BBH} is likely a fast merger ($\lesssim 100$ Myr) while the wide outer orbit merges in longer timescales ($\sim$Gyrs, Figure\,\ref{fig:totalMassProperties}).  
This hierarchical nature leads to a cutoff in the delay time distribution at early times and a possible pile up at higher redshifts, which can be probed with third generation \ac{GW} detectors \citep{LIGO20173G}. It is further supported by the fact that lower metallicity environments are believed to dominate at higher redshifts.

\section{Conclusions}
We investigated configurations of isolated massive stellar triples that lead to sequential \ac{BBH} mergers.
We find that triples with \ac{CHE} inner binaries are good candidates for sequential mergers.
Our model predicts that \ac{GW} sources with one \ac{BH} in the mass gap and $\chi_{\rm{eff}}> 0.1$, can be of sequential merger origin.
We highlight two classes of triples that lead to \acp{BBH} in the mass gap.
The first one has a tertiary \ac{BH} above the mass gap and $0.1 \lesssim \chi_{\rm{eff}} \lesssim 0.27$ (see red sub-region A in Figure \ref{fig:massGap} and Figure \ref{fig:totalMassProperties}).
The second one has a tertiary \ac{BH} below the mass gap and $0.38 \lesssim \chi_{\rm{eff}} \lesssim 0.58$ (see blue sub-region B in Figure \ref{fig:massGap} and Figure \ref{fig:totalMassProperties}).
We suggest \event\ is of triple origin and belongs to the second class.
The masses and spins of mass gap event \eventGap\ are inconsistent with our model, which further supports a cluster origin. 

From a broader point of view, we outlined a new outcome of the evolution of massive stellar triples from a proof-of-principle study. 
To improve upon the predictions made here, several processes should be considered; the effects of the uncertain initial orbital configurations of stellar triples, which may also lead to non-negligible three-body dynamical effects, the non-trivial problem of mass transfer in triples, \ac{GW} recoil kicks from the inner \ac{BBH} merger, and their combined effects on population statistics. 
Higher-order multiplicity in massive stars is crucial to understanding the most energetic astronomical phenomena in the Universe. 

\acknowledgments
The authors thank David R. Aguilera-Dena, Chris Belczynski, Giacomo Fragione, Ryosuke Hirai, Ilya Mandel, Chris Moore, Ataru Tanikawa, and Team COMPAS for useful discussions and suggestions.
We thank Teresa Rebagliato for the graphic representation of the formation channels.
A.V-G. and E.R-R. acknowledge funding support by the Danish National Research Foundation (DNRF132). 
E.R-R. is also supported by the Heising-Simons Foundation and NSF (AST-1911206 and AST-1852393).
S.T. acknowledges support from the Netherlands Research Council NWO (VENI 639.041.645 grants).
N.W.C.L. gratefully acknowledges support from the Chilean government via Fondecyt Iniciaci\'on Grant \#11180005.
C.-J.H. acknowledge support of the National Science Foundation, and the LIGO Laboratory.
LIGO was constructed by the California Institute of Technology and Massachusetts Institute of Technology with funding from the National Science Foundation and operates under cooperative agreement PHY-1764464.

%



\software{
\href{https://compas.science/}{COMPAS} v02.11.04 publicly available at GitHub via  \href{https://github.com/TeamCOMPAS/COMPAS}{TeamCOMPAS/COMPAS}. 
Scripts used for this study available in GitHub via \href{https://github.com/avigna/sequential-mergers}{avigna/sequential-mergers}.
}

\providecommand{\noopsort}[1]{}




\appendix
\section{Rapid population synthesis}
\label{app:COMPAS}
We use the publicly available data from \cite{Riley2020CHE} for synthetic \ac{BBH} formation and merger rates.
That study made use of the COMPAS rapid population synthesis code \citep{stevenson2017formation,VignaGomez2018DNSs,Neijssel2019MSSFR}, which is freely available at \url{http://github.com/TeamCOMPAS/COMPAS}.  
The version of COMPAS used for these simulations was v02.11.01a, built specifically for \cite{Riley2020CHE}; functionality in this release was integrated into the public COMPAS code base in v02.11.04.

\cite{Riley2020CHE} performs, for the first time, simultaneous population synthesis of \ac{CHE} and standard evolving binaries. 
We use the orbital distributions of \ac{BBH} mergers from that study as an educated guess for the properties of the inner \ac{BBH} in the sequential merger scenario.
While this works well as a first approach, the initial conditions from \cite{Riley2020CHE} are based on birth distributions from observations of isolated massive binaries \citep{Sana2012}. 
\cite{Riley2020CHE} simulates binaries with metallicities $-4 \leq \log_{10} Z \leq -1.825$.
As there are no binary evolution models available at zero-metallicity Pop III stars, all of our quantitative results from compact binaries come exclusively from \ac{CHE} binaries.

The most relevant conclusions drawn from this data  concerning sequential mergers are:
\begin{itemize}
    \item 
    There are, overall, $\sim 4\times$ more merging \ac{BBH} from standard evolving than from \ac{CHE} binaries.
    The local merger rate ($z=0$), which accounts for star formation history and galaxy mass-metallicity dependence, are 50 and 20 $\rm Gpc^{-3}\ yr^{-1}$ for isolated binaries and the subset of \ac{CHE} binaries respectively. 
    For \ac{CHE} binaries we include massive over-contact binaries. 
    A massive over-contact binary can fill its inner Lagrangian point during the main sequence as long as the outer Lagrangian point is not filled \citep{Marchant2016CHE}.
    \item 
    The yield of merging \ac{BBH} from \ac{CHE} binaries is roughly constant below $\log_{10} Z/\rm{Z_{\odot}} \lesssim -0.5$ (c.f. Figure 6 of \citealt{Riley2020CHE}), with $\rm{Z_{\odot}}=0.0142$.
    Low-metallicity \ac{CHE} stars are more compact than high-metallicity ones; they also experience less mass loss and orbital widening through stellar winds.
    \item 
    The maximum pre-merger total mass of \acp{BBH} for both \ac{CHE} and standard evolving binaries 
    is $M_{\rm{BH,1}}+M_{\rm{BH,2}} \approx 79\ \rm{M_{\odot}}$.
    \item 
    In this COMPAS data set, there are no \ac{BBH} above the mass gap. This holds even when including stars  with initial masses up to $150\ \Msun$, for which the adopted stellar evolution models from \cite{Hurley_2000} are extrapolated to stars above $50\ \Msun$. 
    However, we consider there is a possibility of \acp{BBH} above the mass gap in Nature.
    They could come from standard evolving stars \citep[e.g.][]{Mangiagli2019gap} or \ac{CHE} binaries \citep[e.g.][]{Marchant2016CHE,duBuisson2020CHE}.
    \item 
    We use the \textit{delayed} supernova mechanism prescription from \cite{fryer2012compact} to determine the remnant mass and natal kick distributions. 
    This prescription predicts complete fallback for \ac{BH} progenitors with carbon-oxygen core masses above $11\ \rm M_{\odot}$.
    In our population, more than half of standard evolving stars and all \ac{CHE} stars leading to \acp{BBH} have masses above this threshold.
    Complete fallback has two implications.
    The first one is that the final baryonic mass of the remnant is the same as the pre-supernova mass (Equations 19 and 20 from \citealt{fryer2012compact}) modulo neutrino emission.
    The second one is that there are no natal kicks for heavy \acp{BH} (Equation 21 of \citealt{fryer2012compact}).
\end{itemize}

We emphasize that a full population synthesis of massive stellar triples will help us understand better the evolution of sequential mergers, their delay time distribution and constrain their rates.
We hope future software developments and observations will make this sort of study possible in the near future.

\setcounter{figure}{4}  
\section{Radiated mass during \ac{BBH} mergers}
\label{app:rad}
For a \ac{BBH}, $f_{\rm{rad}}$ is the amount of energy radiated away in the form of gravitational waves during the coalescence, expressed as a fraction of the total mass of the \ac{BBH}.
While the magnitude of $f_{\rm{rad}}$ is independent of the \ac{BBH} total mass, it does depend on the binary mass ratio $q_{\rm{in}} = M_{\rm{BH,2}}/M_{\rm{BH,1}}$ (with $M_{\rm{BH,1}} \geq M_{\rm{BH,2}}$) and the \ac{BH} spin configuration.
More specifically, it depends on $\chi^{L^{||}}_{1,2} \equiv \chi_{1,2}\cos(\theta_{1,2})$, with $0 \leq \chi_{1,2} \leq 1$ being the \ac{BH} spin magnitude and $\theta_{1,2}$ the zenith angle between the spin and orbital angular momenta at the time of merger for each \ac{BH}.
$f_{\rm{rad}}$ is typically approximated by making use of fitting formulae that are based on numerical relativity simulations of \acp{BBH}.

\begin{figure}
\includegraphics[width=\textwidth]{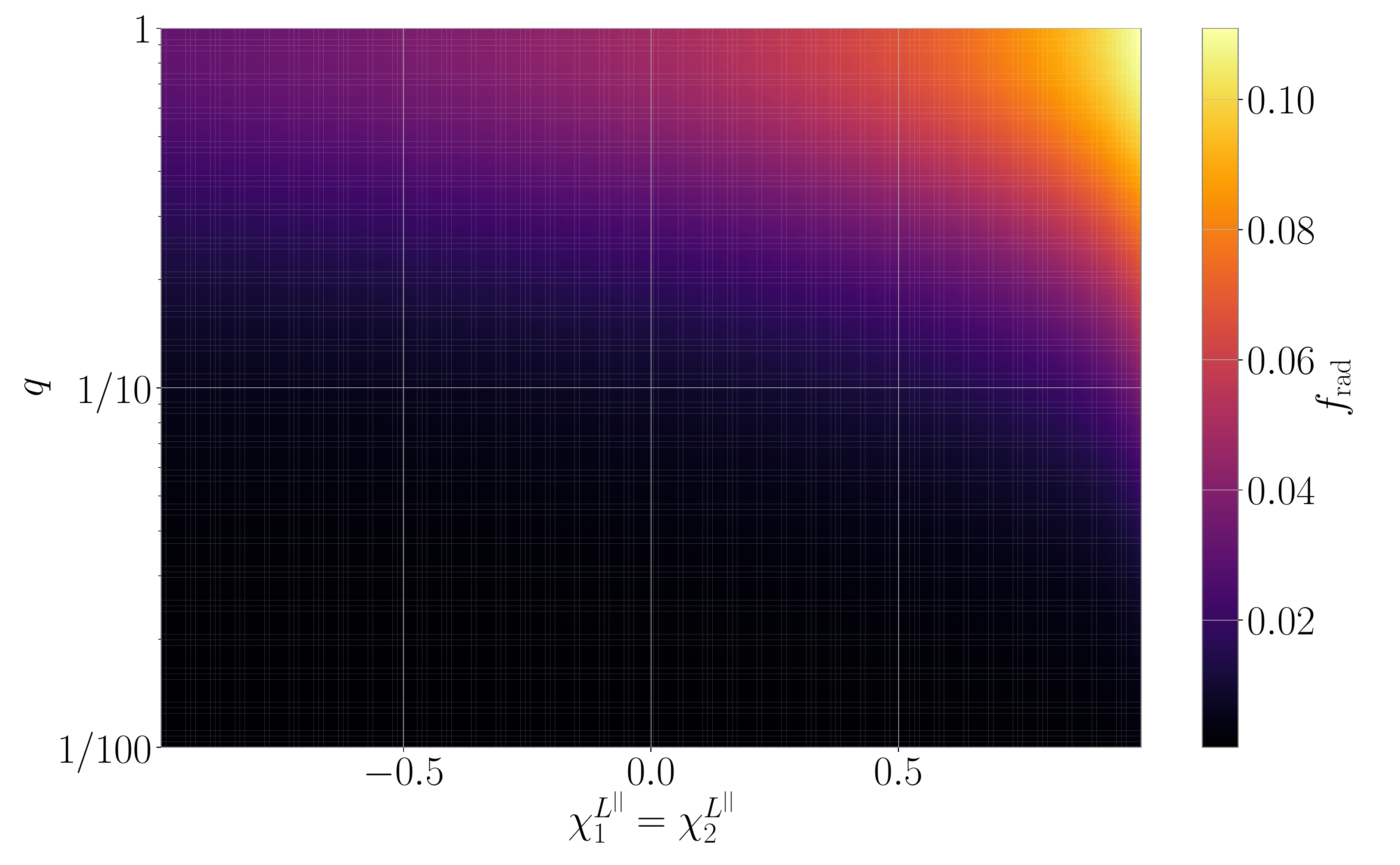}
\centering
\caption{
Radiated mass fraction, $f_{\rm{rad}}$, from a \ac{BBH} merger.
For simplicity, we present the case for equal spins aligned with the orbital angular momentum vector $\hat{L}_{\rm{N}}$.
In the Letter we focus on the non-rotating scenario with $\chi^{L^{||}}_{1,2}=0$.
We use this to estimate the GW Blaauw kick of the inner \ac{BBH} merger in Section\,\ref{sec:method}.
}
\label{fig:radMass}
\end{figure}

In this study, we use the prescription from Equation 28 of~\cite{2017PhRvD..95f4024J} as implemented in LALSuite \citep{lalsuite} and show $f_{\rm{rad}}$ for a selection of \ac{BH} spins and mass ratios in Figure~\ref{fig:radMass}.
For simplicity, and following our assumptions, we have restricted the figure to binaries with equal $\chi^{L^{||}}_{1,2}$ values. 
We use this prescription to estimate the final masses from both the inner \ac{BBH} merger (Figure \ref{fig:totalMass} and \ref{fig:massGap})  and from sequential mergers.

\section{Redshift estimate}
In order to test the validity of the sequential merger scenario for \cite{GWTC12019}, we focus on the expected delay times for our scenario. The delay time is the time it takes a system to evolve from the zero-age main-sequence to the sequential \ac{BBH} merger, and follows from our models. 
We will convert it into a redshift in the following way.
We use \texttt{astropy’s cosmology} module to convert between redshift and lookback time using \texttt{astropy’s lookback\_time} \citep{astropy:2013, astropy:2018}.
We assume a flat $\rm \Lambda CDM$ cosmology with $\rm \textit{H}_{0} = 67.90\ km\ s^{-1}\ Mpc^{-1}$ and $\Omega_{\rm{m}}=0.3065$ following \citealt{Planck2016}. 
With this setup, for a delay time of $\sim 5.1\ \rm Gyr$, we estimate a redshift $z\approx 0.49$.

\section{Intermediate Mass Ratio Inspirals}
In Figure \ref{fig:massGap} we mark the region of sequential mergers with masses $M\ge 124\ \rm{M_{\odot}}$ and mass ratios $q<0.2$ as being \acp{IMRI}. 
This type of systems is degenerate with \acp{IMRI} of binary origin.
\acp{IMRI} have primary masses of e.g., $M_{\rm{BH,1}}\ge 100\ \rm{M_{\odot}}$ and mass ratios of e.g., $q<0.1$ \citep[see, e.g.,][]{Haster2016IMRIs,Haster2016IMRIsb}.
In Section \ref{subsec:IMRIs} we briefly discuss that they are usually associated to cluster origin, but we do not rule out isolated origin. 
In Table \ref{tab:summary} we present the properties of \acp{IMRI} from red sub-region A and yellow sub-region C.

\begin{figure}
\includegraphics[width=0.5\textwidth]{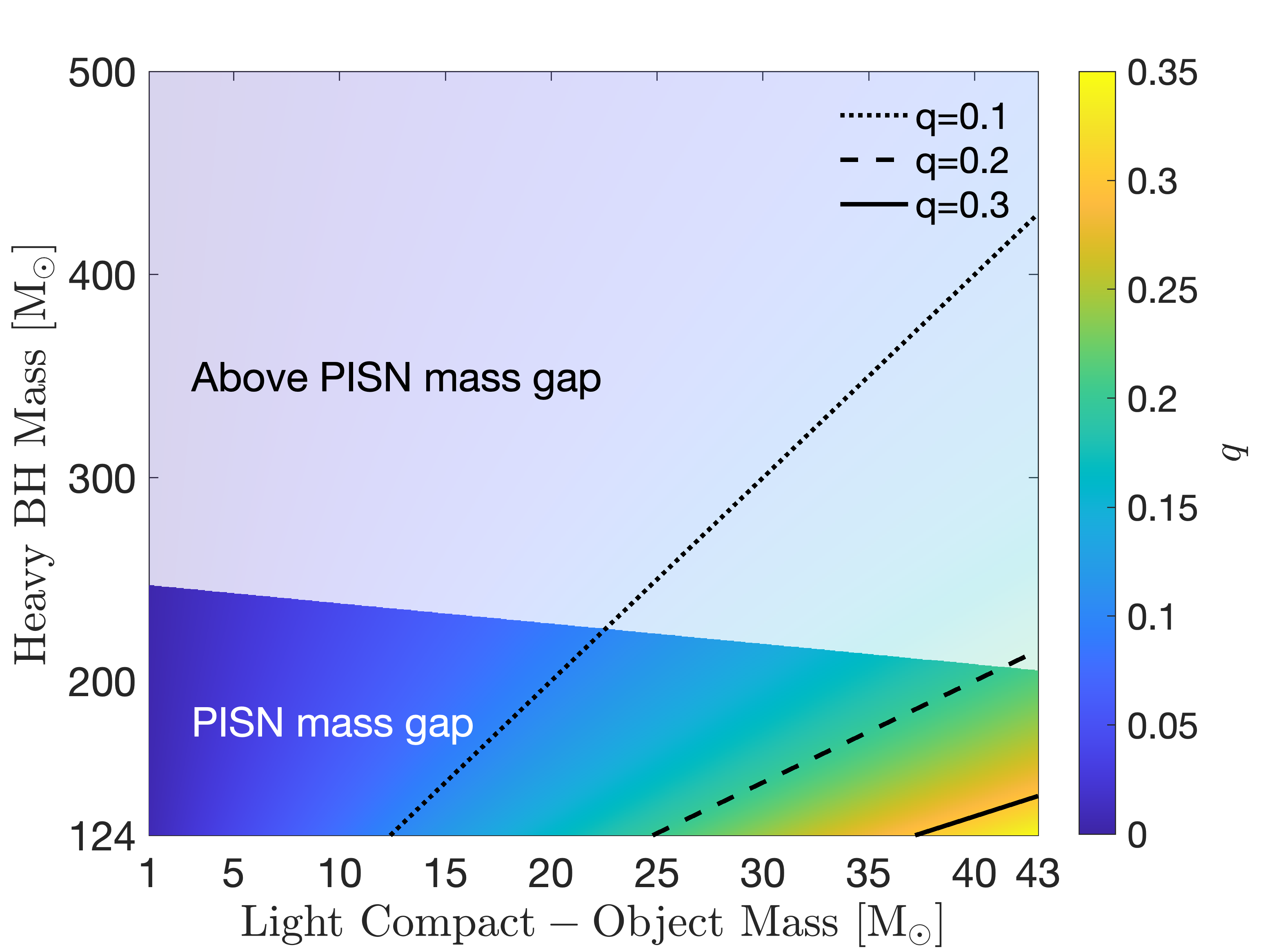}
\includegraphics[width=0.5\textwidth]{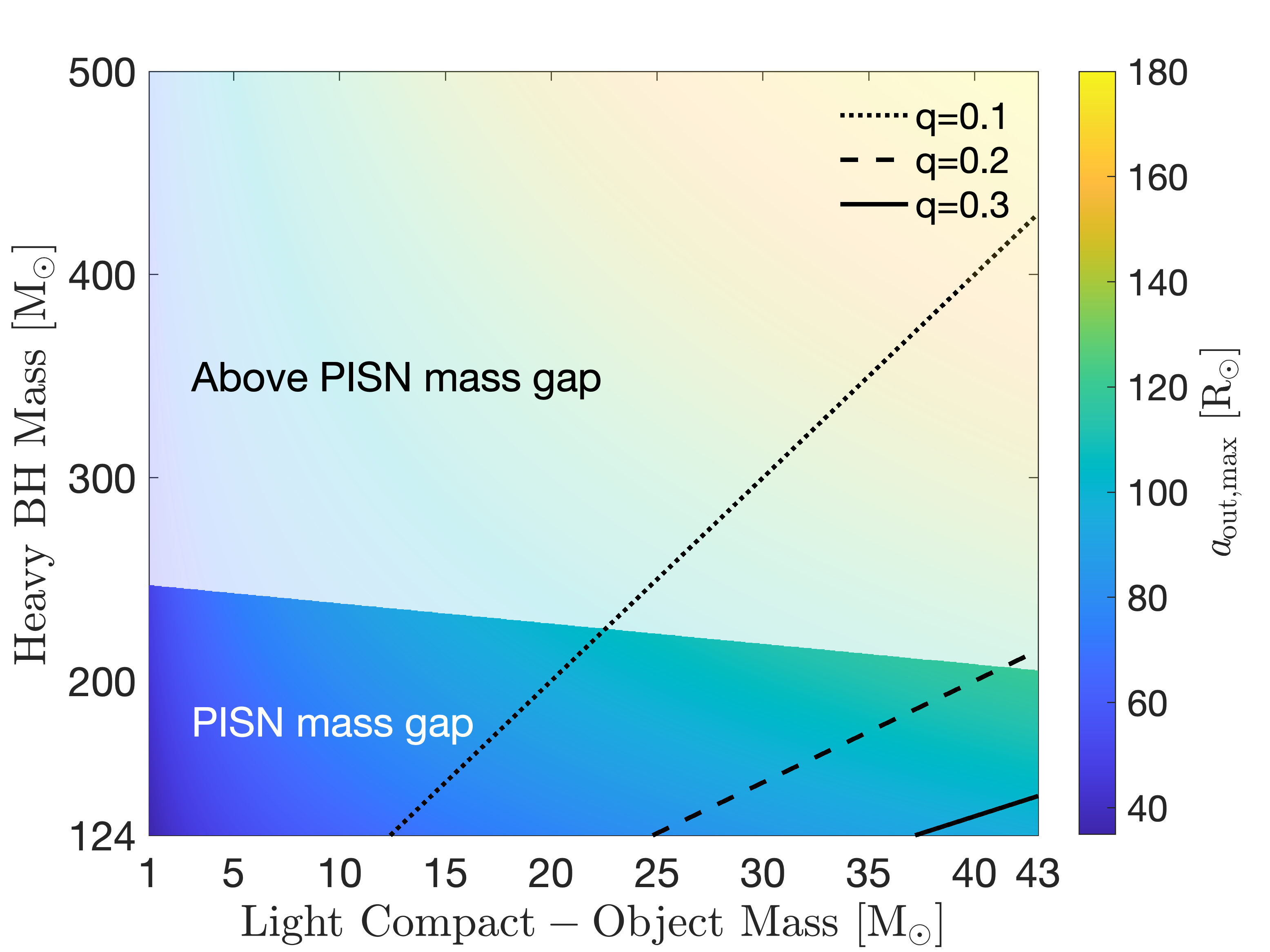}
\caption{
\ac{IMRI} systems can populate the mass gap as binaries in the total mass parameter space, but if their component masses are properly resolved they can be excluded.
The transparency shows the region outside the mass gap.
This plot is similar to the red region from Figure \ref{fig:totalMass} in the limit where $M_{\rm BH,2} \ll M_{\rm BH,1}$. 
}
\label{fig:IMRI}
\end{figure}

In this Appendix we briefly expand our analysis of sequential mergers leading to systems that are degenerate in mass with \acp{IMRI}.
Figure \ref{fig:IMRI} considers double compact objects with primary masses $124 \leq M_{\rm{BH,1}}/\rm{M_{\odot}} \leq 500$ and light compact-object masses $1 \leq M_{\rm{light}}/\rm{M_{\odot}} \leq 43$.
In this case, light compact-objects include both neutron stars and black holes. 
It is not dependent of the maximum mass of neutron stars nor the existence or absence of a mass gap between neutron stars and black holes.
These limits constrain the mass ratio to $0 \lesssim q \lesssim 0.35$.
To date GW190814 is the gravitational-wave source with the most extreme mass ratio of $q = M_{\rm{light}}/M_{\rm{BH,1}} \approx 2.6/23 \approx 0.11$ \citep{Abbott2020GW190814}. 
However, the individual masses of GW190814 are well below those of typical \acp{IMRI}.

For an \ac{IMRI} the inference of the effective spin might not be enough to classify their origin.
For a sequential merger, if the tertiary is the most massive component then the effective spin is dominated by this massive non-rotating \ac{BH}, i.e. $\chi_{\rm{eff}}\approx 0$.
This value would be similar to that of a non-rotating binary.
On the other hand, if the tertiary is the least massive component, that effective spin is  $\chi_{\rm{eff}}\approx 0.7$.
This is visually summarized in Figure \ref{fig:totalMassProperties} and quantitatively in Table \ref{tab:summary}.

More detailed studies of the spin properties of \ac{BBH} mergers from isolated binary, isolated triple and cluster origin would help in the classification of future detections of \acp{IMRI}.
Additionally, population studies might be helpful in constraining the merger rates from different formation channels. 

\section{Sequential mergers rate estimate}
\label{app:rate}
In order to give an upper limit in the local ($z=0$) rate of sequential mergers ($\mathcal{R}$) we make a simple estimate in the form
\begin{equation}
    \label{eq:app_rates}
    \mathcal{R}=\mathcal{R}_{\rm{BBH,in}}\times (f_{\rm{triple}}/f_{\rm{binary}}) \times f_{\rm{separation}}.
\end{equation}
We estimate the \ac{BBH} merger rate $\mathcal{R}_{\rm{BBH,in}}=50\ \rm Gpc^{-3}\ yr^{-1}$ based on the isolated binary evolution calculations of \cite{Riley2020CHE}, which accounts for both compact and standard evolving stars.
This optimistic rate includes sequential mergers with total mass below the mass gap.
\cite{Riley2020CHE} assumes that a fraction $f_{\rm{binary}}=0.7$ of massive star systems are  binaries.
The factor $f_{\rm{triple}}=0.35$ describes the fraction of massive star systems that are stellar triples \citep{MoeDiStefano2017}. 

The factor $f_{\rm{separation}}$ accounts for the fraction of systems in which the outer orbital can merge within a Hubble time due to gravitational-wave emission. 
To estimate $f_{\rm{separation}}$ we assume the log of the outer birth orbital separation is distributed uniformly  $p(a_{\rm{out}})\propto a_{\rm{out}}^{-1}$ \citep{opik1924statistical,Abt1983}.
We then estimate the fraction of systems of interest in the form:
\begin{equation}
\label{eq:f_separation}
    f_{\rm{separation}} = 
    \dfrac{\int_{a_{\rm{out-SM,min}}}^{a_{\rm{out-SM,max}}} a^{-1}da}{\int_{a_{\rm{out,min}}}^{a_{\rm{out,max}}} a^{-1}da} =
    \dfrac{\int_{35\ \rm{R_{\odot}}}^{135\ \rm{R_{\odot}}} a^{-1}da}{\int_{28\ \rm{R_{\odot}}}^{2\times 10^6\ \rm{R_{\odot}}} a^{-1}da} \approx 0.12,
\end{equation}
where $a_{\rm{out}}$ is the outer separation of stable triples, and $a_{\rm{out-SM}}$ is the outer separation of potential sequential mergers.
The lower limit on $a_{\rm{out}}$ is given by the smallest value for a stable triple with an inner binary separation $a_{\rm{in}}\approx 10\ \rm{R_{\odot}}$, $(a_{\rm{out}}/a_{\rm{in}})_{\rm{crit}}=2.8$, and therefore $a_{\rm out,min}=28\ \rm{R_{\odot}}$.
The maximum outer separation at birth we consider is $\rm \textit{a}_{out,max}\approx 10^4\ AU \approx 2\times10^6\ R_{\odot}$ \citep{MoeDiStefano2017}. 
For the outer separation of potential sequential mergers, following the results of Table \ref{tab:summary} as shown in top right panel of Figure \ref{fig:totalMassProperties}, the limits are $35 \lesssim a_{\rm{out-SM}}/R_{\odot} \lesssim 135$.
These limits in $a_{\rm{out-SM}}$ assume that the separation does not drastically change from the zero-age main sequence until the inner \ac{BBH} merger.
For some triples, this assumption will not necessarily hold (Section \ref{sec:disc}).
For when it holds, we expect the distribution of merger times to be $p(t)\propto t^{-1}$, as expected from gravitational-wave dominated binaries with a flat-in-the-log distribution at \ac{BBH} formation. 
This assumption also neglects initially wider tertiary stars which are stripped and become potential sequential mergers (Section \ref{subsubsec:red}), which would increase the values of $a_{\rm{out-SM,max}}$, $f_{\rm{separation}}$ and ultimately the rate.
The assumption of different orbital initial distributions would naturally also affect the rates.
We choose these assumptions for an order-of-magnitude estimate and leave a more thorough analysis of the orbital parameter space and the distribution of merger times for a future study.

After substituting all of the estimated parameters in Equation \ref{eq:app_rates}, we constrains the upper limit to $\rm \mathcal{R} < 3\ Gpc^{-3}\ yr^{-1}$ for sequential mergers.

\section{More details on the evolution of \event}
\label{app:dynamics}

As a proof of concept of the evolution towards sequential mergers, we have simulated the evolution of the proposed progenitor of \event\ with two independent codes: the rapid binary population synthesis code COMPAS and the triple evolution code \texttt{TRES} \citep{Toonen2016TrES,Toonen2020TRES}.
We use these codes to model the evolution from the zero-age main sequence to the inner \ac{BBH} merger.

\begin{figure}
\includegraphics[width=0.5\textwidth]{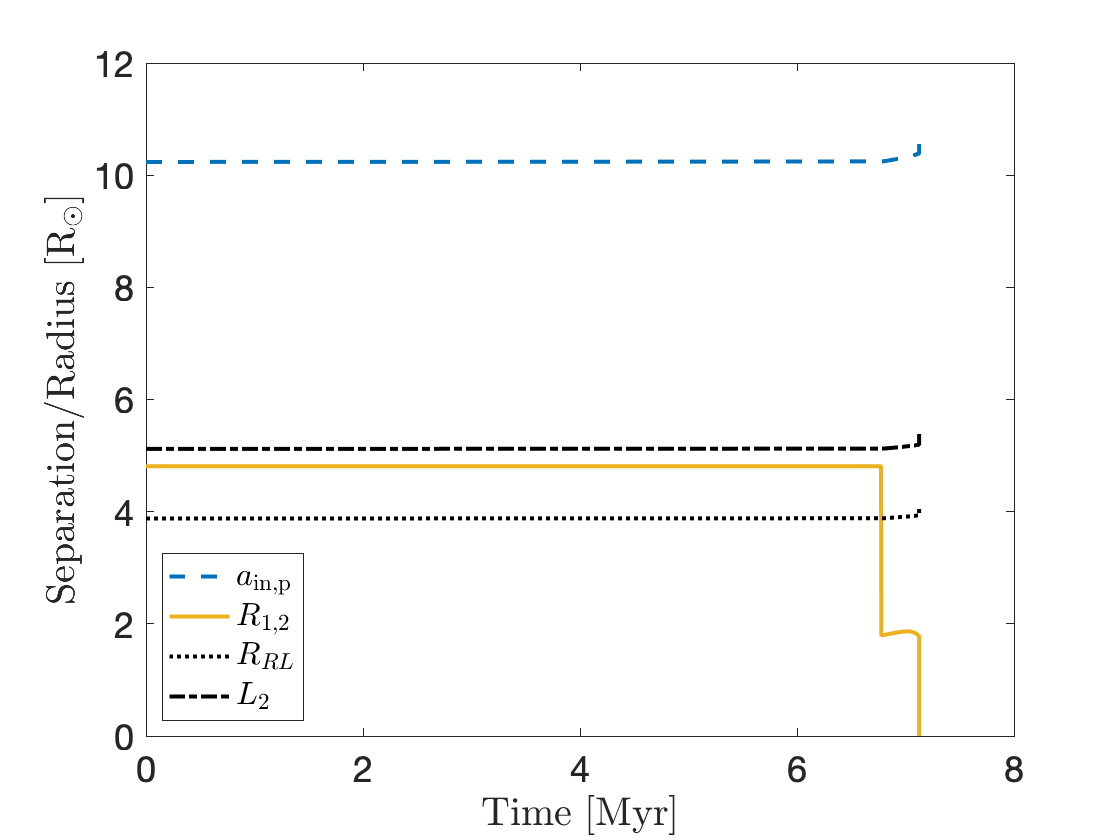}
\includegraphics[width=0.5\textwidth]{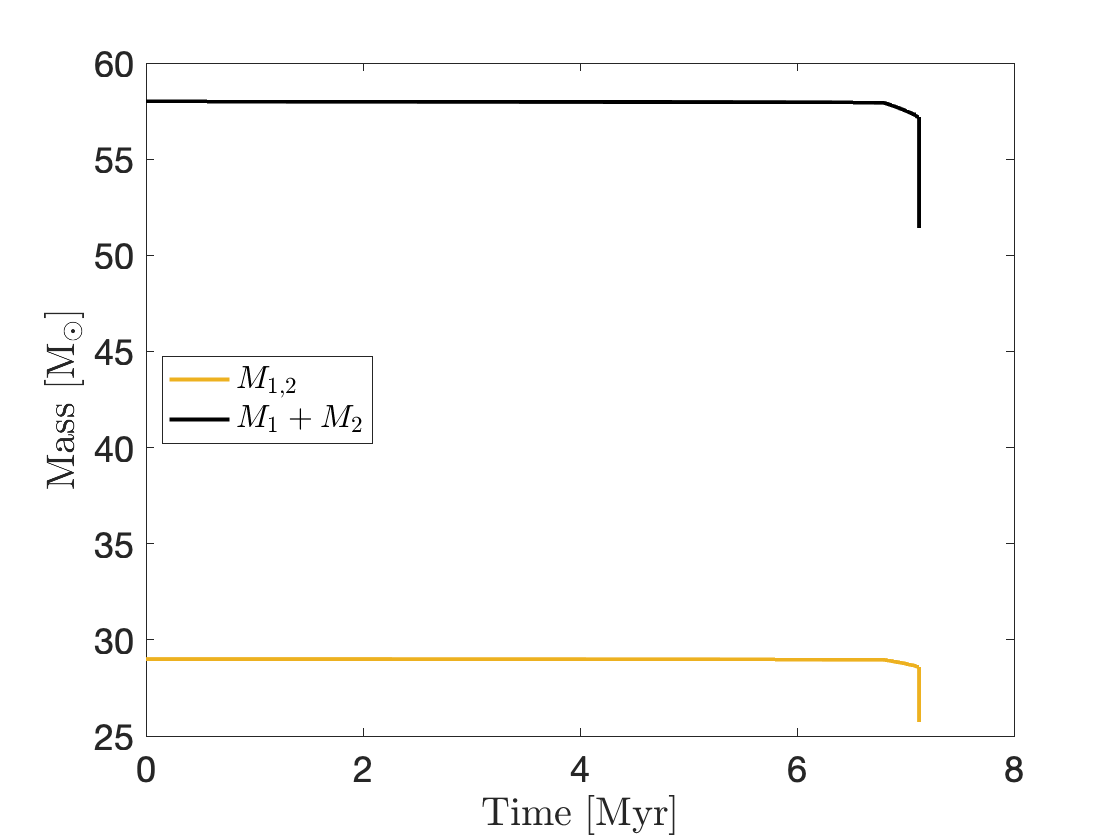}
\centering
\caption{
COMPAS time evolution, from zero-age main sequence to \ac{BBH} formation, of a \ac{CHE} binary at $Z=10^{-4}$.
Top: separation (dashed blue), radius (solid yellow), Roche lobe (black dotted) and second Lagrangian point (black dot-dashed).
Bottom: individual masses (solid yellow) and total mass (solid black).
}
\label{fig:popSynth}
\end{figure}

We use COMPAS as described in Appendix \ref{app:COMPAS} to explore a binary system that is representative of the \ac{CHE} inner binary for a \event-like system.
We follow the evolution from the zero-age main sequence until \ac{BBH} formation.
This system consists initially of a circular \ac{CHE} binary with $M_1=M_2=29\ \rm{M_{\odot}}$, $R_1=R_2=4.8\ \rm{R_{\odot}}$ and $a=10.2\ \rm{R_{\odot}}$.
The separation, masses, and radial time evolution is shown in Figure \ref{fig:popSynth}.
\ac{CHE} are expected to contract throughout their lifetimes \citep[see, e.g.,][]{Aguilera-Dena2018ApJ...858..115A}; however, the current implementation of them in COMPAS assumes a fixed radius during the main sequence.
The low metallicity results in negligible mass loss and the orbit barely changes throughout the full evolution.
There are two milestones in the evolution.
The first one is the evolution after hydrogen depletion ($\approx 6.8$ Myr), when each component becomes a $1.8\ \rm{R_{\odot}}$ naked helium star.
The final one is the failed supernova ($\approx 7.1$ Myr), which reduces the mass by 10\% (Section \ref{app:COMPAS}), increases the separation to $a=11.6\ \rm{R_{\odot}}$ and the eccentricity to $e=0.07$.
This \ac{BBH} merges in $\approx 76$ Myr \citep{peters1964gravitational}.

\begin{figure}
\includegraphics[width=0.5\textwidth]{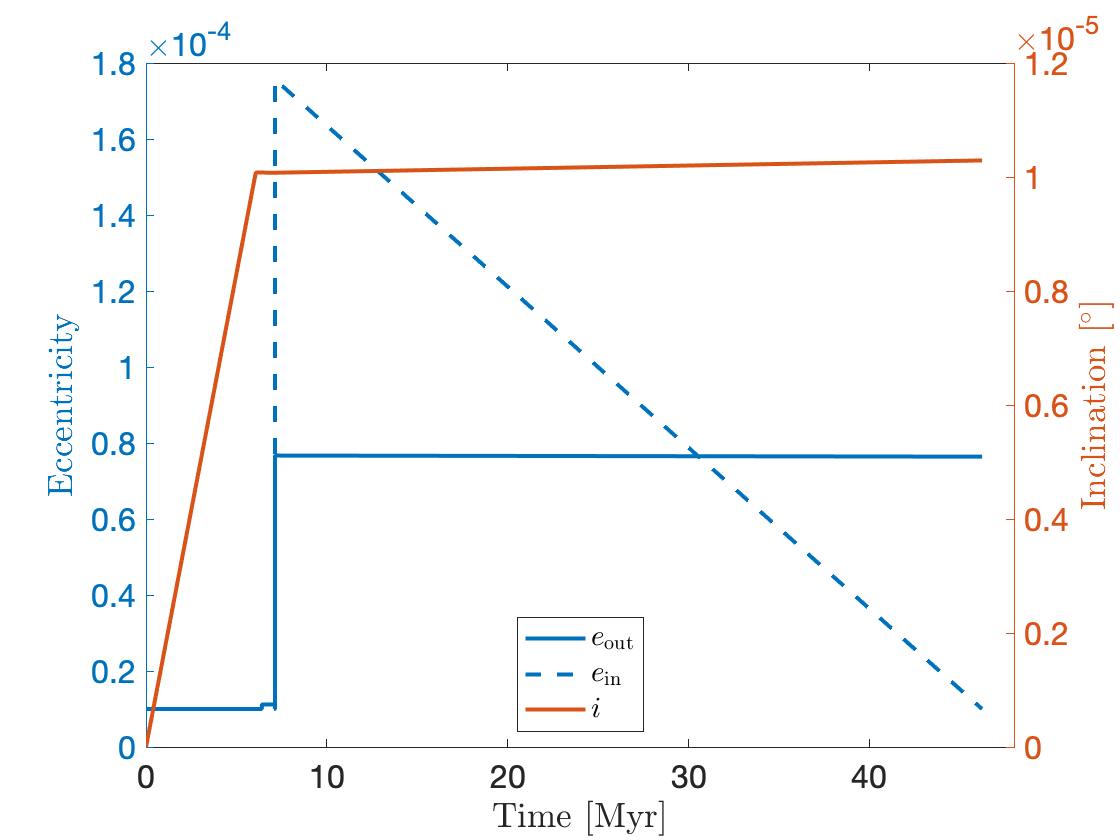}
\includegraphics[width=0.5\textwidth]{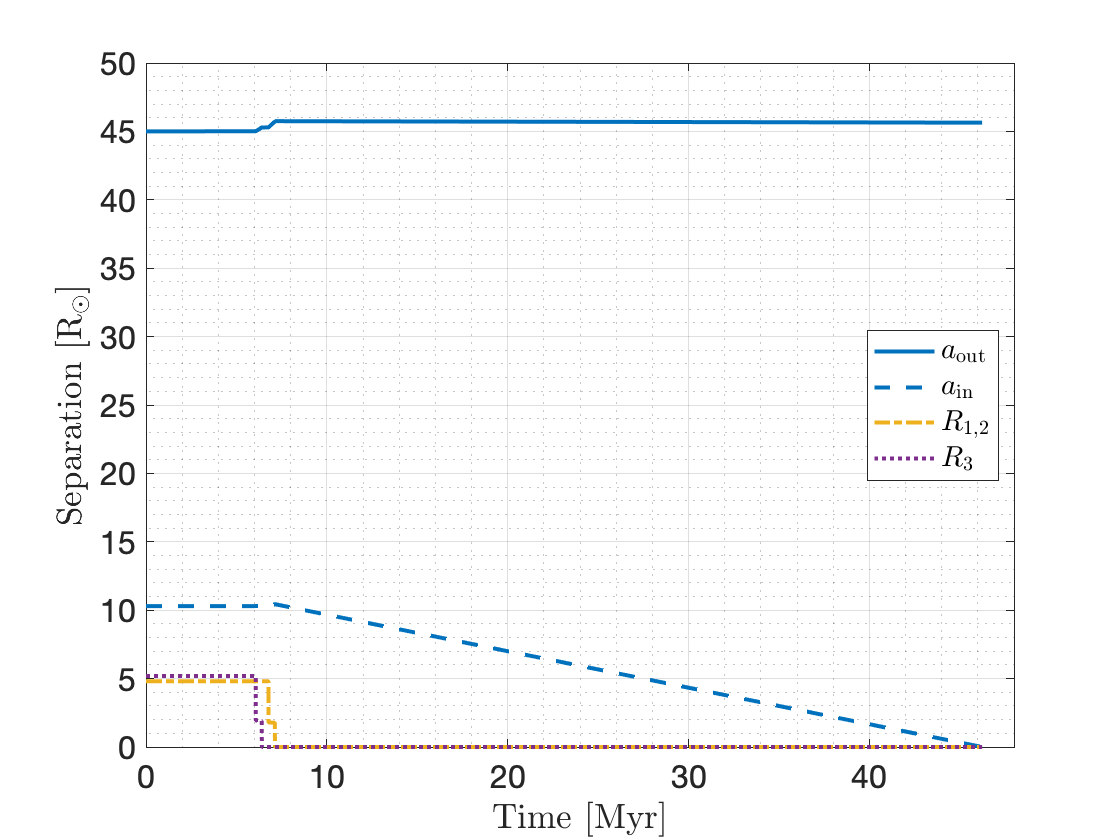}
\includegraphics[width=0.5\textwidth]{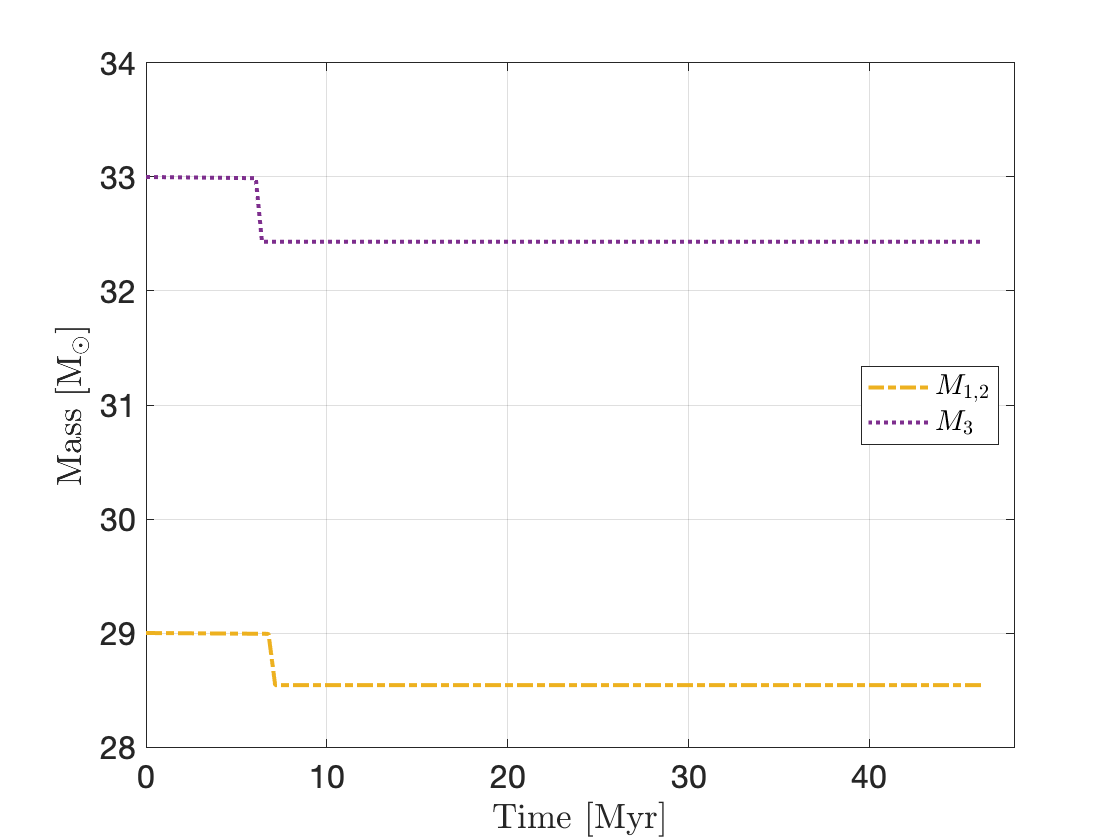}
\centering
\caption{
\texttt{TRES} time evolution, from zero-age main sequence to inner \ac{BBH} merger, of a triple leading to a \event-like system (Section \ref{subsec:event}).
Top: outer eccentricity (solid blue), inner eccentricity (dashed blue) and relative inclination (solid orange).
Middle: outer separation (solid blue), inner separation (dashed blue), radius of the stars in the inner binary (dot-dashed yellow) and radius of the tertiary star (dotted purple).
Bottom: mass of the stars in the inner binary (dot-dashed yellow) and mass of the tertiary star (dotted purple).
}
\label{fig:TRES}
\end{figure}

Additionally, we modified the triple evolution code \texttt{TRES} \citep{Toonen2016TrES,Toonen2020TRES} to model the evolution, from the zero-age main sequence until the \ac{BBH} coalescence of a compact inner binary of a triple system.
The inner binary consists of \ac{CHE} system with $M_1=M_2=29\ \rm{M_{\odot}}$, $R_1=R_2=4.8\ \rm{R_{\odot}}$, $a_{\rm{in}}=10.3\ \rm{R_{\odot}}$ and $e_{\rm{in}}=10^{-5}$ (this eccentricity is the lower limit allowed in \texttt{TRES} for numerical reasons).
Similarly to COMPAS, the radial evolution of all compact stars throughout the main sequence is kept constant.
The tertiary is also assumed to be a compact star with $M_3=33\ \rm{M_{\odot}}$, and $R_3=5.2\ \rm{R_{\odot}}$ in an orbit of $a_{\rm{out}}=45\ \rm{R_{\odot}}$, $e_{\rm{out}}=10^{-5}$, and relative inclination $i=0.0$.
The complete orbital evolution is shown in Figure \ref{fig:TRES}.
There are four milestones in the evolution of this triple system.
The first one is the formation of an outer $33.4\ \rm{M_{\odot}}$ \ac{BH} at $\approx 6$ Myr.
The second one is the evolution after hydrogen depletion ($\approx 6.8$ Myr), when each component of the inner binary becomes a $1.8\ \rm{R_{\odot}}$ naked helium star.
The third one is the supernovae ($\approx 7.1$ Myr) of the stars in the inner binary which result in two \acp{BH} of $M_{\rm{BH,1}}=M_{\rm{BH,2}}=28.5\ \rm{M_{\odot}}$, $a_{\rm{in}}=10.4\ \rm{R_{\odot}}$, $e_{\rm{in}}\approx 2\times 10^{-4}$, $e_{\rm{out}}\approx 0.8\times 10^{-4}$ and $i\approx 10^{-5}$.
The final milestone is the gravitational-wave evolution from the inner binary; the inner \ac{BBH} merges at $\approx 46$ Myr.
The triple remains approximately circular and co-planar throughout the evolution, validating our assumptions at the moment of triple \ac{BH} formation and throughout the sequential merger.

In summary, we used two independent codes to test the validity of our assumptions in Section \ref{sec:method}.
With COMPAS we validate the compact inner binary evolution.
We use \texttt{TRES} to do a dynamical calculation to corroborate that, following our initial conditions, there are no major orbital changes which would modify our assumptions during both \ac{BBH} mergers.
Finally, we show that our numerical results are consistent with our semi-analytical formalism.

\end{document}